\documentstyle[12pt,aas2pp4]{article}
\begin{document}
\title{X-ray Constraints on Accretion and Starburst Processes in 
Galactic Nuclei I. Spectral Results}
\author{A. Ptak\altaffilmark{1}, P. Serlemitsos, T. 
Yaqoob\altaffilmark{2}, R. Mushotzky}
\affil{NASA GSFC, Code 662, Greenbelt, MD 20771}
\altaffiltext{1}{Present address: Carnegie Mellon University, 
Department of Physics, Pittsburgh, PA 15213}
\altaffiltext{2}{Also with USRA}
\slugcomment{Accepted Jul. 30, 1998}
\begin{abstract}
The results of the analysis of 0.4-10.0 keV {\it ASCA} spectral 
analysis of a sample of low-luminosity AGN (LLAGN; M51, NGC 3147, NGC 4258), 
low-ionization nuclear emission line regions (LINERs; NGC 3079, NGC 
3310, NGC 3998, NGC 4579, NGC 4594) and starburst galaxies (M82, NGC 253, 
NGC 3628 and NGC 6946) are presented. In spite of the heterogeneous 
optical classifications of these galaxies, the X-ray spectra are fit 
well by a ``canonical'' model consisting of an optically-thin 
Raymond-Smith plasma ``soft'' component with $T \sim 7 \times 10^{6}$ K 
and a ``hard'' component that can be modeled by either a power-law 
with a photon index $\Gamma \sim 1.7$ or a thermal bremsstrahlung with 
$T \sim 6 \times 10^{7}$ K.  The soft component absorption is 
typically less than $10^{21} \rm \ cm^{-2}$ while the hard component 
is typically absorbed by an additional column on the order of $10^{22} 
\rm \ cm^{-2}$.  The soft-component 0.4-10 keV instrinsic 
luminosities tend to be 
on the order $10^{39-40} \rm \ ergs \ s^{-1}$ while the hard-component 
luminosities tend to be on the order of $10^{40-41} \rm \ ergs \ 
s^{-1}$.  The abundances inferred from the fits to the soft-component 
are significantly sub-solar.  The Fe abundance can be measured 
independently of the other elemental abundances (dominated by 
$\alpha$-process elements) in M51, M82, NGC 253, and NGC 4258.  In 
these galaxies the Fe abundance relative to $\alpha$-process elements 
is also (statistically) significantly sub-solar.  There is some indication
(at a low-statisical signficance) that the abundance properties starburst
emission from starburst galaxies differs from the starburst emission from
low-luminosity AGN.  However, these results on
abundances are model-dependent. Significant Fe-K line emission is 
observed in M51, M82, NGC 3147, NGC 4258 and NGC 4579.  An analysis of 
the short-term variability properties was given in 
(\cite{ptak98a}) and detailed interpretation of these results will be 
given in Paper II (\cite{ptak98b}).
\end{abstract}
\keywords{galaxies: abundances --- galaxies:active --- galaxies: ISM ---
galaxies: starburst --- accretion --- ISM: abundances --- ISM: kinematics and
dynamics --- X-rays: galaxies --- X-rays: ISM}

\twocolumn
\section{Introduction}
Observations by the imaging X-ray satellite {\it Einstein} 
showed that ``normal'' spiral galaxies tend to have X-ray luminosities 
on the order of $10^{38-41} \rm \ ergs \ s^{-1}$ and spectra 
consistent with the superposition of (typical) X-ray binary spectra
 (see \cite{fab89} for a review).
In some cases, extended X-ray emission and multiple extra-nuclear 
point-sources were detected (c.f., M82 in \cite{wsg84} and 
\cite{fab88}; NGC 253 in \cite{ft84} and \cite{fab88}; M51 in 
\cite{pfft85}).  However, the high spectral resolution of
{\it ASCA} (see \cite{tih94}) is necessary to provide further
insight into the nature of
the X-ray emission.  Early results from the {\it ROSAT} All-Sky Survey (\cite{bol92}) found 
that some non-AGN spiral galaxies may be as bright as $L_{0.1-2.2\ \rm keV}
\sim 
10^{42} \rm \ ergs \ s^{-1}$ and suggested that these high X-ray 
luminosities may be due to very powerful starburst activity.  An optical
follow-up to this study 
showed that RASS galaxies with $L_{\rm \ 0.1-2.2 keV} \gtrsim 
10^{42} \rm \ ergs \ s^{-1}$ tended to be low-luminosity AGN (LLAGN; 
\cite{mor94}; \cite{mor96}) and the {\it ASCA} spectra of one of these
galaxies, NGC 3147, revealed X-ray properties consistent with Seyfert 2 
galaxies (see \cite{ptak96}).

In this paper we present the results of {\it ASCA} spectral analysis of
a sample of ``low-activity'' galaxies consisting of LLAGN, LINERs, and
starbursts.  This sample is listed in Table 1, which also lists the 
Galactic column densities and the distances to each galaxy, and Table 2 
gives a log of the {\it ASCA} observations.  The sample
is based on 
galaxies with public and proprietary {\it ASCA} data available to us 
as of the summer of 1996 and is
not complete in any sense.  However, this study will hopefully give 
some insight into the nature of the X-ray emission beyond that which 
might be expected from ``normal'' galaxies.  In cases where there is 
strong evidence for the presence of a low-luminosity AGN, this analysis will 
demonstrate the extent to which the X-ray properties of these 
galaxies are simply an extension of the X-ray properties of normal 
Seyferts to lower luminosities.  As discussed in Ptak et al. (1998a), the
variability behavior of these low-activity galaxies differs significantly
from the trends of normal AGN.  Specifically, much less short-term variability
is observed than expected.  This suggests a different mode of accretion,
and this may also have spectral implications as well (see Paper II for a
full discussion). 

\footnotesize
\begin{deluxetable}{cccccc}
\tablehead{
\colhead{Galaxy} & \colhead{R.A., Decl.} &
\colhead{Type*} & \colhead{$N_{H, Gal.}$} &
\colhead{Distance} & \colhead{ASCA PSF**} \nl
& \colhead{(J2000)} & & \colhead{($10^{20} \rm \ cm^{-2}$)} &
\colhead{(Mpc)} & \colhead{(kpc)}} 
\startdata
NGC 253 & 00h47'33.1, -25d7'18 & H & 1.3$^{a}$
& 2.5$^{b}$ & 2.2 \nl
NGC 3079 & 10 01 57.8, 55 40 47 & S2 & 0.8$^{a}$
& 28.8$^{c}$ & 25 \nl
NGC 3147 & 10 16 53.2, 73 24 04 & S2 & 2.5$^{d}$
& 56.4$^{c}$ & 49 \nl
NGC 3310 & 10 38 45.9, 53 30 12 & H & 1.4$^{j}$ &
19.4$^{c}$ & 17 \nl
NGC 3628 & 11 20 17.0, 13 35 20 & T2 & 2.2$^{j}$ &
14.9$^{c}$ & 13 \nl
NGC 3998 & 11 57 56.1, 55 27 13 & L1.9 & 1.0$^{j}$ &
24.3$^{f}$ & 21 \nl
NGC 4258 & 12 18 57.5, 47 18 14 & S1.9 & 1.2$^{j}$ &
10.2$^{g}$ & 8.9 \nl
NGC 4579 & 12 37 43.5, 11 49 05 & S1.9/L1.9 & 3.1$^{j}$ &
36.1$^{c}$ & 32 \nl
NGC 4594 & 12 39 58.8, -11 37 28 & L2 & 3.8$^{j}$ &
22.6$^{c}$ & 20 \nl
NGC 6946 & 20 34 52.3, 60 09 14 & H & 20-50$^{a,h}$ &
8.3$^{g}$ & 7.2 \nl
M51 & 13 29 52.4, 47 11 54 & H & 1.3$^{a}$ &
14.0$^{c}$ & 12 \nl
M82 & 09 55 52.2, 69 40 47 & H & 4.3$^{a}$ &
3.6$^{i}$ & 3.1 \nl
\tablerefs{a) Fabbiano et al. (1992), b) de Vaucouleurs (1963), c) 
Soifer et al. (1987), Stark et al (1992), e) Hartmann \& Burton (1995),
f) de Vaucouleurs et al. (1991), g) Tully (1988), h) Burstein \& Heiles
(1984), i) Freedman et al. (1994), j) Murphy et al. (1996)}
\tablenotetext{*}{Optical classification given in Ho et al. (1997a) (the
optical classification for NGC 253 is from NED).  H=HII region/starburst
galaxy, T = transition starburst/LINER galaxy,
L = LINER galaxy, S = Seyfert galaxy.}
\tablenotetext{**}{Distance corresponding to 3' $\sim$ the {\it ASCA}
half-power diameter.}
\enddata
\end{deluxetable}

\setcounter{table}{1}
\begin{deluxetable}{ccccccc}
\tablecaption{{\it ASCA} Observation Log}
\tablehead{
\colhead{Galaxy} & \colhead{Number\tablenotemark{a}} & \colhead{Date} &
\colhead{CCD Mode\tablenotemark{b}} &
\colhead{Exposure (ks)\tablenotemark{c}}
& \colhead{SIS Region Size}}
\startdata
NGC 253 & 60038000 & Jun. 12, 1993 & 4 & 33.5, 33.6, 36.2, 36.2 & 6 \nl
NGC 3079 & 60000000 & May 9, 1993 & 1 & 41.0, 41.1, 40.8, 40.2 & 4 \nl
NGC 3147 & 60040000 & Sep.29, 1993 & 4 & 22.4, 23.0, 38.7, 39.0 & 4 \nl
NGC 3310 & 61013000 & Apr. 17, 1994 & 1 &17.7, 15.3, 17.9, 17.9 & 4\nl
NGC 3310 & 61013010 & Nov. 13, 1994 & 2 & 10.2, 9.9, 10.3, 10.3 & 4\nl
NGC 3310 & 61013020 & Apr. 11, 1995 & 1 & 10.7, 10.6, 11.3, 11.3 & 
4\nl
NGC 3628 & 61015000 & Dec. 12, 1993 & 2 & 17.5, 20.2, 23.9, 23.9 & 3 \nl
NGC 3998 & 71048000 & May 10, 1994 & 2 & 40.8, 38.6, 40.0, 39.9 & 4 \nl
NGC 4258 & 60018000 & May 15, 1993 & 4 & 33.0, 34.1, 40.1, 40.1 & 6 \nl
NGC 4579 & 73063000 & Jun. 25, 1995 & 2 & 36.0, 34.0, 34.6, 34.6 & 4 \nl
NGC 4594 & 61014000 & Jan. 20, 1994 & 2 & 19.6, 18.2, 20.1, 20.1 & 4 \nl
NGC 6946 & 60039000 & May 31, 1993 & 4 & 20.2, 20.5, 27.0, 27.0 & 4 \nl
M51 & 60017000 & May 11, 1993 & 4 & 32.4, 33.1, 38.1, 38.1 & 6 \nl
M51 & 15020100 & May 4, 1994 & 4 & 16.9, 11.2, 42.5, 42.5 & 6 \nl
M82 & 60001000 & Apr. 19, 1993 & 4 & 18.1, 18.2, 27.8, 27.3 & 6 \nl
\tablenotetext{a}{Refence number for this observation in the {\it ASCA} 
archival database}
\tablenotetext{b}{Number of active CCDs}
\tablenotetext{c}{Exposure times for SIS0, SIS1, GIS2, and GIS3}
\enddata
\end{deluxetable}

\setcounter{table}{2}
\begin{deluxetable}{cccc}
\tablecaption{{\it ROSAT} Observation Log}
\tablehead{
\colhead{Galaxy} & \colhead{Number\tablenotemark{a}}
& \colhead{Dates} & 
\colhead{Exposure (ks)}
}
\startdata
NGC 253 & rp600087a00 & Dec. 25-31, 1991 & 11.6 \nl
NGC 253 & rp600087a01 & Jun. 3-5, 1992 & 11.2 \nl
NGC 3079 & rp700319n00 & Nov. 14-15, 1991 & 16.3 \nl
NGC 3147 & wp600589n00 & Oct. 16-22, 1991 & 8.5 \nl
NGC 3310 & rp600156n00 & Nov. 17-18, 1991 & 8.0 \nl
NGC 3628 & rp700010n00 & Nov. 23-6, 1991 & 13.9 \nl
NGC 3998 & rp700055 & May 22-24, 1991 & 55.9 \nl
NGC 4258 & rp600186n00 & May 9, 1992 & 5.3 \nl
NGC 4258 & wp600546n00 & Nov. 11-15, 1993 & 25.4 \nl
NGC 4579 & rp700056n00 & Dec. 15-16, 1991 & 8.9 \nl
NGC 4594 & rp600258n00 & Jul. 15-19, 1992 & 10.6 \nl
NGC 6946 & rp600272n00 & Jun. 16-21, 1992 & 35.6 \nl
M51 & rp600158n00 & Nov. 28 - Dec. 13, 1991 & 22.2 \nl
M82 & rp600110a00 & Mar. 27-29, 1991 & 10.6 \nl
M82 & rp600110a01 & Oct. 15-16, 1991 & 10.7 \nl
M82 & wp600576n00 & Sep. 29 - Oct. 3, 1993 & 17.0 \nl
\tablenotetext{a}{Refence number for this observation in the {\it ROSAT} 
archival database}
\enddata
\end{deluxetable}
\normalsize

{\it ROSAT} PSPC spectra were also analyzed along with the {\it ASCA} 
data, with emphasis on improving the determination of the absorbing 
column along the line of sight.  While the spectral resolution of the 
PSPC is relatively poor, the sensitivity of {\it ROSAT} to photons in the 
0.1-0.3 keV range makes it much more sensitive to columns on the 
order of $10^{19-20} \rm \ cm^{-2}$ than {\it ASCA}.  A log of the PSPC
observations 
discussed in this paper is given in Table 3.

\section{Summary of Previous X-ray Observations}
 Dahlem, Weaver, \& Heckman (1998; hereafter DWH) 
discuss spatial analyisis of the {\it ROSAT} PSPC and HRI data and spectral
analysis of the {\it ASCA} and {\it ROSAT} pspc data for a sample of starburst
galaxies, including NGC 253, NGC 3079,
NGC 3628, and M82 which are discussed here.  DWH found evidence for
three-component fits to the {\it ASCA} and {\it ROSAT} PSPC spectra from these
galaxies, which are discussed in \S6.7.

\subsection{NGC 253}
NGC 253 was observed by pointed observations with both the {\it Einstein} HRI 
(\cite{ft84}) and IPC (\cite{fab88}).  The HRI 
observation shows that there were $\sim$ 8 extra-nuclear X-ray sources, 
with X-ray luminosities in the range of $1-5 \times 10^{38} \ \rm ergs 
\ s^{-1}$.  The 
nucleus is extended over $\sim 12"$, occupying a similar region as the 
nuclear radio and IR emission but not elongated along the major axis 
as these components are.  The estimated luminosity of the nucleus was 
$\sim 3 \times 10^{39} \ \rm ergs \ s^{-1}$.  Of course, these
luminosity estimates are 
sensitive to assumptions concerning the source spectra, particularly 
the assumed amount of absorption.  The IPC observation, which is more 
appropriate for studying diffuse flux, showed that the X-ray radial 
profile is steeper than the optical (blue) profile along the major 
axis and agrees better with the radio profile.  The X-ray emission 
along the minor axis, associated with superwind outflow, extends to $\sim 
9$ kpc, further than the optical or radio flux.  The X-ray radial 
profile along the minor axis was consistent with a density 
distribution $\rho \sim r^{-2}$ as would be expected for a 
freely-expanding wind.  The energy resolution of the IPC was rather poor 
($\Delta E/E \sim 100\%$) but
showed that the spectrum of NGC 253 is consistent with a $\sim 5$ keV 
bremsstrahlung model with absorption on the order of the Galactic 
column density, $\sim 2 \times 10^{20} \rm \ cm^{-2}$.
Fabbiano (1988) estimates the halo gas density, mass and radiative lifetime at 
$1.7 \times 10^{-3} f^{-1/2} \rm \ cm^{-3}$, $2.2 \times 10^{7} 
f^{1/2} M_{\odot}$, and $2.1 \times 10^{9} f^{1/2}$ yr, where $f$ is 
the volume filling factor (see Paper II for a detailed discussion).

\cite{O90} found that {\it Ginga} detected NGC 253 with a 2-10 
keV flux of $\sim 7.0 \times 10^{-12} \rm \ ergs \ cm^{-2} \ s^{-1}$ 
($L_{2-10 \rm \ keV} \sim 9.7 \times 10^{39} \rm \ ergs \ s^{-1}$). \cite{bh94}
report the detection of 
NGC 253 by the {\it Compton Gamma-Ray Observatory (CGRO)} at a flux level 
about an order of magnitude above the extrapolated Ginga flux.  This 
detection was interpreted by \cite{gr95} as 
inverse-Compton scattered IR flux and microwave cosmic background 
flux.  The {\it BBXRT} 0.3-10.0 keV spectrum of NGC 253 was fit well with a 
double Raymond-Smith plasma model with temperatures of $\sim$ 0.6 and 
6 keV (\cite{petre93}).
{\it ROSAT} PSPC data showed that the 0.1-0.4 keV flux 
is most prominent in the halo while the 0.5-2.4 keV flux is most 
prominent in the disk, suggesting that the halo emission is 
significantly cooler than the disk emission (\cite{piet95}).  \cite{read97}
found 7 sources in addition to the {\it Einstein} sources.
The PSPC data show that
several of the sources are hard (kT $>$ several keV) and are therefore likely
to be X-ray binaries rather than supernova remnants.  \cite{read97} also found
that $\sim 74\%$ of the flux in the PSPC bandpass appears to be ``diffuse'',
although the high value of $\chi^2_{\nu}$ for a thermal fit to the diffuse
component suggests that a significant fraction of the diffuse flux may be due
to unresolved point sources.
Analysis 
of the {\it ASCA} data was presented in Ptak et al. (1997) and is discussed 
in the remainder of this paper. 

\subsection{NGC 3079}
NGC 3079 was detected by the {\it Einstein} IPC at the 3.2s level 
(\cite{ffz82}), with an estimated flux of
$3.7 \times 10^{-13} \rm \ ergs \ cm^{-2} \ s{-1}$. \cite{rmf94} found that
the {\it ROSAT} PSPC data from NGC 3079 is dominated by a nuclear point source 
but that X-ray emission is detected up to 2.5' from the nucleus.  
Above 0.5 keV the X-ray emission is slightly elongated along the 
(optical and radio) major axis.  Below 0.5 keV there is a ``soft patch''
1-2' northeast of the nucleus.  The best-fitting spectral model for 
the galaxy was a Raymond-Smith plasma plus power-law model with kT $\sim$ 
0.4 keV and a photon index $\Gamma$ of $\sim$ 1.6.
\cite{read97} found that $\sim 69\%$ of the flux from NGC 3079 is diffuse in
the {\it ROSAT} bandpass. The diffuse PSPC flux is fit well by a thermal model
(with kT $\sim 0.5$ keV and abundances $\sim 0.03 \times$ solar).

\subsection{NGC 3147}
NGC 3147 was not detected significantly by {\it Einstein}, although emission 
at the 3$\sigma$ level was detected by {\it HEAO-I} in the vicinity of NGC
3147 (\cite{rgp95}).  Analysis of the {\it ASCA} data was 
presented in Ptak et al. (1996).  The authors found that the spectrum appeared
to be absorbed by a column $< 10^{21} \rm \ cm^{-2}$, which indicated that
either the nucleus was truely unobscurred, or that the X-ray flux was
scattered into the line of sight.  In addition, the detection of a 
large equivalent width Fe line and the optical classification of the
source as a Seyfert 2 strongly indicates that the X-ray flux from NGC 3147
is indeed scattered into the line of sight.

\subsection{NGC 3628}
Although NGC 3628 is a starburst galaxy, it has exhibited dramatic X-ray 
variability. The 0.1-2.0 keV X-ray flux detected by the {\it Einstein} IPC 
$\sim 8 \times 10^{-13} \ \rm ergs \ cm^{-2} \ s^{-1}$ in 1979 and the X-ray
flux in this 
bandpass detected by the {\it ROSAT} PSPC in 1991 was at approximately the 
same level, but the {\it ASCA} flux was at a level of $3 \times 
10^{-13} \ \rm ergs \ cm^{-2} \ s^{-1}$ in 1993 and a 
HRI observation in 1994 resulted in an upper limit of $\sim 0.5 \times
10^{-13} \ \rm ergs \ cm^{-2} \ s^{-1}$ (\cite{dhf95}).
(Note that the {\it Einstein} and 
{\it ASCA} flux estimates in this bandpass require some extrapolation since 
their sensitivity only extends down to $\sim$ 0.4 keV.) This 
variability obviously shows that the nucleus is dominated by a 
point-source, most likely a micro-AGN or an X-ray binary.  The X-ray 
luminosity of $\sim 1 \times 10^{40} \ \rm ergs \ s^{-1}$, if sub-Eddington, 
implies that the mass of any 
accreting object would have to be $> 75 M_{\odot}$.  Diffuse X-ray emission 
associated with the halo of NGC 3628 is discussed in Dahlem et al. (1996) and
is evidence for a superwind similar to the ones observed in M82 and NGC 253.
The {\it ASCA} spectrum was observed to be very flat, leading \cite{yaq95} to 
suggest that if a large population of galaxies with a similar 
spectrum is detected, then these galaxies may contribute 
significantly to the X-ray background.

\section{NGC 3310}
NGC 3310 was detected by {\it Einstein} with a 0.5-4.5 keV flux of
$1.1 \times 10^{-12} \ \rm ergs \ cm^{-2} \ s^{-1}$ and luminosity of 
$4.9 \times 10^{41} \rm \ ergs \ s^{-1}$ (\cite{ft92}).  A detailed analysis
of the {\it ASCA} and {\it ROSAT} data of NGC 3310 in \cite{zgw98} showed that
the soft emission is extended over $\sim 3$ kpc and that, as with other
starbursts such as NGC 253 and M82, the broadband (0.1-10.0 keV) X-ray
spectrum consists of at least two components: a kT $\sim 0.8$ keV soft
component and a hard component fit well with a power-law with a photon index
of $\sim 1.4$ or a thermal component with kT $\sim 15$ keV.

\section{NGC 3998}
Awaki et al.  (1991) found that the Ginga flux measured for NGC 3998 
was about 3 times larger (extrapolated) than the 0.4-3.5 keV {\it Einstein}
flux ($5.0 \times 10^{-12} \rm \ ergs \ cm^{-2} \ s^{-1}$) given in
\cite{dw85}.  
A thermal model fit to NGC 3998's 2-20 keV spectrum could be rejected while 
a power-law moel with a slope of $\sim$ 2 provided a good fit.  
\cite{rmf94} found that the best-fit to the {\it ROSAT} PSPC spectrum for NGC 
3998 was a broken power-law with slopes of 1.3 and 2.1 and a break 
energy of $\sim$ 0.8 keV.  No extended emission was found in the PSPC 
data.

\subsection{NGC 4258}
{\it Einstein} detected NGC 4258 with a luminosity of $5.9 \times 10^{39} \rm
\ ergs \ s^{-1}$
(Fabbiano et al. 1992).  A {\it ROSAT} PSPC observation showed that the 
extranuclear X-ray flux in NGC 4258 is dominated by jets (\cite{cwd95}),
and that the X-ray emission is due to 
shock-heating along the jet.  Makishima et al. (1994) found that 
{\it ASCA}'s 0.4-10.0 keV spectra of NGC 4258 exhibited at least two 
components: hot gas associated with the jet and a highly-absorbed 
power-law associated with a Seyfert 2 nucleus.  NGC 4258 is one of the 
few galaxies with a well-determined blackhole mass (\cite{greenhill95}), 
allowing its Eddington luminosity to be determined.  Lasota et al.
(1995) modeled the low luminosity of the AGN in NGC 4258 with an 
advection-dominated accretion flow, however \cite{nm95} 
modeled the accretion flow in a ``normal,'' optically-thick (albeit 
warped) accretion disk.

\subsection{NGC 4579}
NGC 4579 was detected by {\it Einstein} with a 0.5-4.5 keV luminosity of 
$\sim 4.7 \times 10^{40} \rm \ ergs \ s-^{1}$
(\cite{gaw92}).  Reichert et al. (1994)
found that the {\it ROSAT} PSPC flux is unresolved with an 
upper limit on the source extent of $\sim$ 5".  The PSPC spectrum is 
best fit with a Raymond-Smith plus power-law model with kT $\sim$ 0.5 keV 
and a photon index $\sim$ 1.9.  A detailed analysis of the {\it ASCA} 
data presented in this paper is given in Terashima et al. (1998b).

\subsection{NGC 4594}
An {\it Einstein} IPC observation of NGC 4594 detected an 0.5-4.5 keV X-ray 
luminosity of $\sim 8.7 \times 10^{40} \ \rm ergs \ s^{-1}$
(\cite{gaw92}).  An analysis of ROSAT PSPC and HRI data by \cite{fj97}
 found that the nucleus is dominated by a point-like source with a 
0.1-2.4 keV luminosity of $\sim 3.5 \times 10^{40} \ \rm ergs \ s^{-1}$,
most likely due 
to a low-luminosity AGN. Given the high blackhole mass estimated by 
\cite{kr95} of $\sim 5 \times 10^{8} \ M_{\odot}$, this AGN 
emission appears to be highly sub-Eddington ($L_{Edd} \sim 
6 \times 10^{46} \rm \ ergs \ s^{-1}$).  There are 3 non-nuclear point-like
sources with a (total) $L_X \sim 2.1 \times 10^{40} \rm \ ergs \ s^{-1}$ and
unresolved emission with $L_X \sim 3 \times 10^{40} \rm \ ergs \ s^{-1}$ that
may be due hot ISM or unresolved point-sources 
(such as X-ray binaries).

\subsection{NGC 6946}
The {\it Einstein} IPC detected two sources in NGC 6946, with a composite 
spectrum consistent with a thermal model with kT $\sim$ 0.8 keV 
(\cite{ft87}).  A {\it ROSAT} PSPC observation resolved the 
X-ray emission from NGC 6946 into 8 sources, 3 of which constitute the 
IPC nuclear source (\cite{sch94b}).  The brightest source is not the 
nucleus but rather a bright source $\sim$ 2' to the north of the 
nucleus with $L_X \sim 3 \times 10^{39} \rm \ ergs \ s^{-1}$ that is
interpreted as being 
a bright supernova remnant (\cite{sch94a}).  The remaining sources 
each have a luminosity of $\sim 10^{38} \rm \ ergs \ s^{-1}$ and diffuse
flux that 
traces the spiral-arm structure is detected with a total luminosity of 
$\sim 10^{39} \rm \ ergs \ s^{-1}$.
The diffuse emission is more consistent with 
hot ISM than unresolved point sources.  The point sources are 
consistent with individual X-ray binaries or supernovae.

\subsection{M51}
The {\it Einstein} HRI observation of M51 detected 3 unresolved sources in 
addition to nuclear and diffuse emission associated with M51 and M51's 
companion, NGC 5195 (\cite{pfft85}).  
Palumbo et al.  found that the point-sources had estimated X-ray 
luminosities on the order of $\sim 10^{39} \rm \ ergs \ s^{-1}$.
M51's nucleus is 
extended over a $\sim$ 24" region, with a point-source contribution of 
less than 20\% to the total luminosity of $\sim$
$8 \times 10^{39} \rm \ ergs \ s^{-1}$.
The X-ray radial profile followed the blue light 
radio profile more closely than the spiral arm structure, indicating 
that a large fraction of the X-ray flux is due to disk population 
sources (i.e., X-ray binaries).  A correlation between the X-ray 
profile and the non-thermal radio profile also suggests an association 
between X-ray emission and the relativistic electron population.  A 
{\it Ginga} observation of M51 detected a 2-10 keV flux and luminosity of 
$6.1 \times 10^{-12} \rm \ ergs \ cm^{-2} \ s^{-1}$ and
$6.7 \times 10^{40} \rm \ ergs \ s^{-1}$ (\cite{ot92})
based on a power-law spectrum with a photon index of $\sim 1.4$.  
A {\it ROSAT} PSPC observation detected eight individual sources in addition 
to nuclear and diffuse emission in M51, with luminosities in the range 
$5-29 \times 10^{39} \rm \ ergs \ s^{-1}$ (\cite{mar95}).
Of these sources, four 
appear to be associated with star-forming regions suggesting that the 
X-ray flux is due to young stellar processes, such as type-II 
supernovae, massive X-ray binaries, or possible "superbubbles" caused 
by multiple supernovae.  A detailed analysis of the PV-phase {\it ASCA} is
given in Terashima et al. (1998b) (note that this paper also includes the
later AO-1 observation of M51).

\subsection{M82}
M82 is one of the most studied galaxies at all wavelengths.  Watson, 
Stanger \& Griffiths (1984) found that the {\it Einstein} HRI data consists 
of up to 8 point-like sources with typical $L_X \sim 5-9 \times 10^{38} 
\rm \ ergs \ s^{-1}$, 
some of which could be identified with nuclear HII regions or SN, 
comprising $\sim 50\%$ of the nuclear flux.  The nuclear region is 
extended over $\sim$ 30" and has $L_X \sim 7.4 \times 10^{39} \rm \ ergs \ 
s^{-1}$.  Halo emission 
that is well-correlated with H$\alpha$ emission is detected out to a radius 
of $\sim 3$ kpc.  Fabbiano (1988) found that the {\it Einstein} IPC
detected the 
X-ray halo out to a radius of $\sim$ 9 kpc with gas density, mass and 
radiative lifetime at $2.1 \times 10^{-3} f^{-1/2} \rm \ cm^{-3}$,
$3.4 \times 10^{7} f^{1/2}$ M$_{\odot}$, and 
$1.7 \times 10^{9} f^{1/2}$ yr, where $f$ is the volume filling 
factor (see Paper II).  As with NGC 253, the X-ray surface brightness 
along the minor axis is consistent with a $\rho \sim r^{-2}$ density 
distribution.

\cite{schaaf89} found that {\it EXOSAT} detected M82 with a 1.4-8.9 
keV luminosity of $3.2 \times 10^{40} \rm \ ergs \ s^{-1}$ out to a radius
of $\sim 6$ kpc.  The 
X-ray emission in this bandpass was fit well by a power-law model with
photon index $\sim 1.8$ or a thermal bremsstrahlung model with kT 
$\sim 9$ keV. 
Schaaf et al.  suggest that most of the X-ray emission is non-thermal 
and is produced by inverse-Compton scattering of IR photons.
\cite{tsuru90} reported a {\it Ginga} 2-10 keV luminosity 
for M82 of $\sim 4 \times 10^{40} \rm \ ergs \ s^{-1}$ and found evidence
for diffuse 
emission extended over $\sim 100$ kpc.  However, it must be kept in mind 
that {\it Ginga} is a non-imaging telescope so source confusion could be 
significant.

{\it ROSAT} HRI observations showed that two of the X-ray sources in M82 
are variable over time scales of years, and one of these, in close 
proximity to the nucleus, is variable on time scales of weeks (
\cite{col94}; \cite{ptak97}).  A detailed analysis of the {\it ROSAT}
HRI data by \cite{bst95} showed that, in addition to the bright point-source 
mentioned above and two other sources with $L_X \sim 10^{38} \rm \ 
ergs \ s^{-1}$, the 
nuclear emission is consistent with an exponential distribution with 
an e-folding length of $\sim 0.27$ kpc.  X-ray emission is detected along 
the minor axis out to a distance of $\sim 6$ kpc and has a surface 
brightness distribution beyond 1.6 kpc of the nucleus consistent with 
a power-law with an exponent of 2-3, again consistent with a 
freely-expanding wind.  However, Strickland et al. (1997) found that 
the halo emission is more consistent with shock-heated clouds than a 
free-flowing wind on the basis of {\it ROSAT} HRI and PSPC data.  
\cite{ml97} found that the {\it ASCA} and {\it ROSAT} PSPC data from M82
are consistent with a power-law with slope of 1.7 and two thermal 
components with temperatures of $\sim 0.3$ and 0.6 keV. Moran \& Lehnert (1997)
claim that the Raymond-Smith models for these thermal components do 
not adequately fit the data, but Ptak et al. (1997) and \cite{tsuru97}
found that acceptable Raymond-Smith fits can 
be found.  \cite{ml97} argue that the hard X-ray emission 
is likely to be due to inverse-Compton scattering of IR photons, while 
Ptak et al. (1997) found that the hard X-ray flux is 
most likely dominated by accreting sources, either blackhole candidate 
binaries or an AGN.  Since these interpretations cannot be resolved
spectrally (in the 2.0-10.0 keV bandpass), high-resolution imaging above 2 keV
(as will be provided by 
{\it AXAF} and {\it XMM}) and/or detection of unambiguous variability above
2 keV will be required to distinguish among them.

\section{Data Analysis Procedures}
A detailed description of the data analysis procedures is given in 
Ptak (1997). Briefly, 
{\it ASCA} consists of two solid-state imaging spectrometers 
(SIS; hereafter S0 and S1) with an approximate bandpass of 0.4-10.0 
keV and two gas imaging spectrometers (GIS; hereafter S2 and 
S3)  with an approximate bandpass of 0.8-10.0 keV.  Times of high
background were 
excluded and hot pixels (in the SIS data) were removed.
The sizes of the SIS source regions and the net exposure times
are listed in Table 2.  A 4' and 6' source region sizes would be appropriate
SIS and GIS (respectively) sizes for a point source, 
however extended emission is clearly evident in some sources, requiring 
larger source regions for the SIS spectra (all of the sources are 
roughly point-like in the case of the GIS and so a 6' region was used).
The background was estimated in two ways.  
First, background counts were extracted from an annulus exterior to 
the source region; hereafter backgrounds estimated in this way are 
referred to as ``local''.  Second, background counts were summed from blank 
sky events lists in the same region as the source counts.  The blank 
sky events lists were obtained from archives at legacy.gsfc.nasa.gov, 
and cleaning criteria consistent with the orbital parameters of each 
observation were applied.  These backgrounds are referred to as 
blank sky backgrounds.  The local backgrounds are most likely more 
robust since below $\sim 1$ keV the SIS blank sky backgrounds are dominated 
by Galactic emission which varies as a function of Galactic position.  
The spectra have been binned to 20 
counts per bin to allow the use of the $\chi^{2}$ statistic.  The data 
from the individual chips of each SIS were combined, however the 
spectra from each detector was fit separately (with the model 
parameters tied except for the overall normalization).  The agreement 
in flux from the individual detectors is discussed in \S 6.3.
Fits were attempted at first to just SIS spectra (which has higher 
spectral resolution than the GIS) and then to the SIS and GIS 
spectra.  The best-fitting {\it ASCA} model was then also simultaneously
fit to any available PSPC spectra (see \S 6.3).  In all of the 
fits, all free parameters other than normalizations were considered 
to be ``interesting'' in determining 90\% confidence intervals (see 
\cite{cash}).  Errors were computed automatically in the course of 
batch processing of the spectra although errors 
should be treated with caution for fits with 
$\chi^{2}_{\nu} \gtrsim 1.5$, where systematics may be important.
Note that since these fits 
have on the order of hundreds of 
degrees of freedom, fits with $\chi^{2}_{\nu} \sim 1.5$ can be 
rejected at a high level of confidence statistically, although again this 
does not incorporate any systematic error (the confidence probability with
which the fits
can be rejected is given in each table in parentheses with the $\chi^2$
values).

\section{Spectral Results}
\subsection{Single-Component Fits}
\begin{figure*}
\plotfiddle{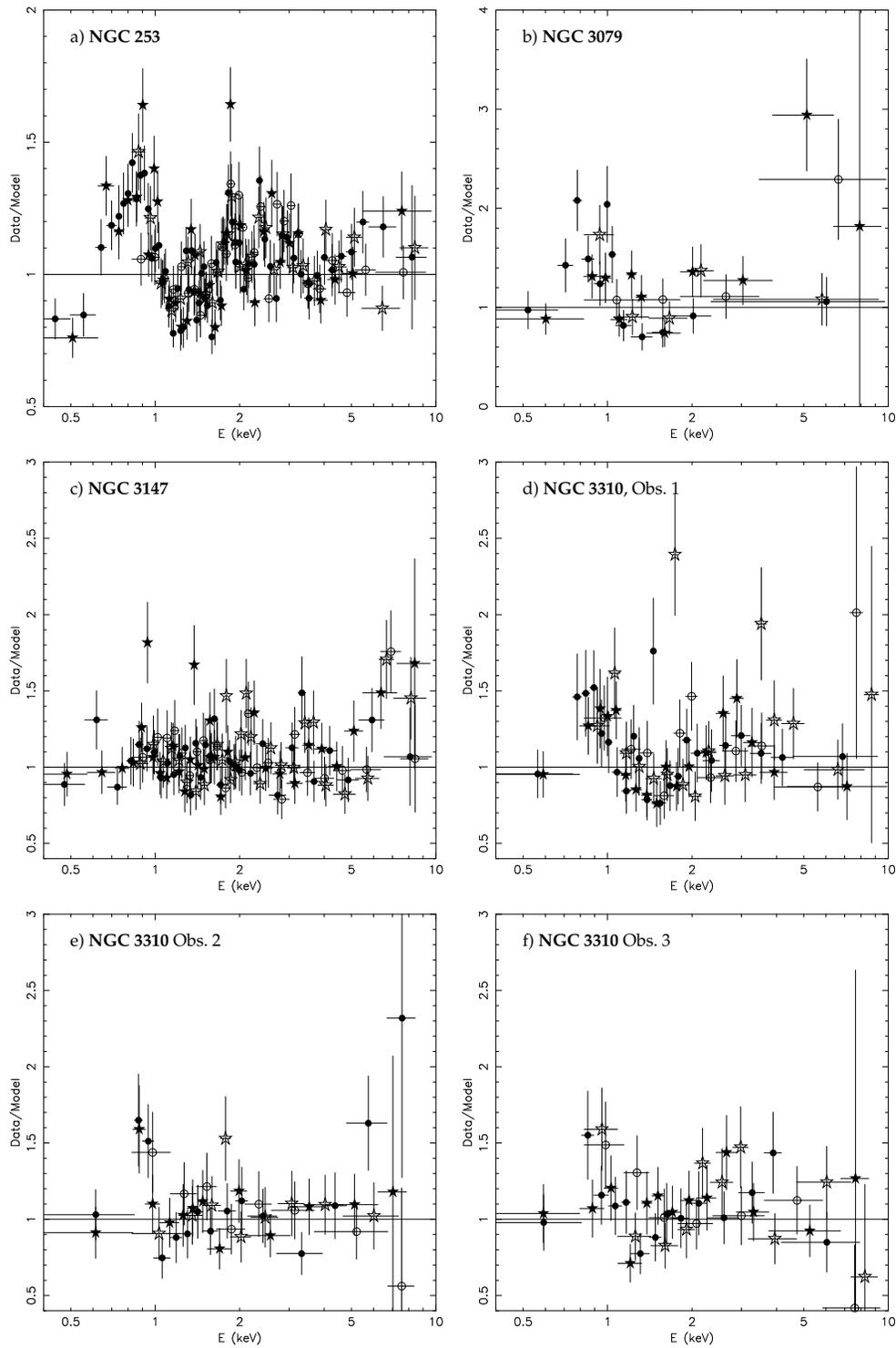}{7in}{0}{100.0}{100.0}{-350}{-170}
\caption{Ratio of data to power-law
model (including neutral absorption) for
each galaxy.  Points show data from S0 (filled circles), S1 (filled stars),
G2 (empty circles) and G3 (empty stars).}
\end{figure*}
\setcounter{figure}{0}
\begin{figure*}
\plotfiddle{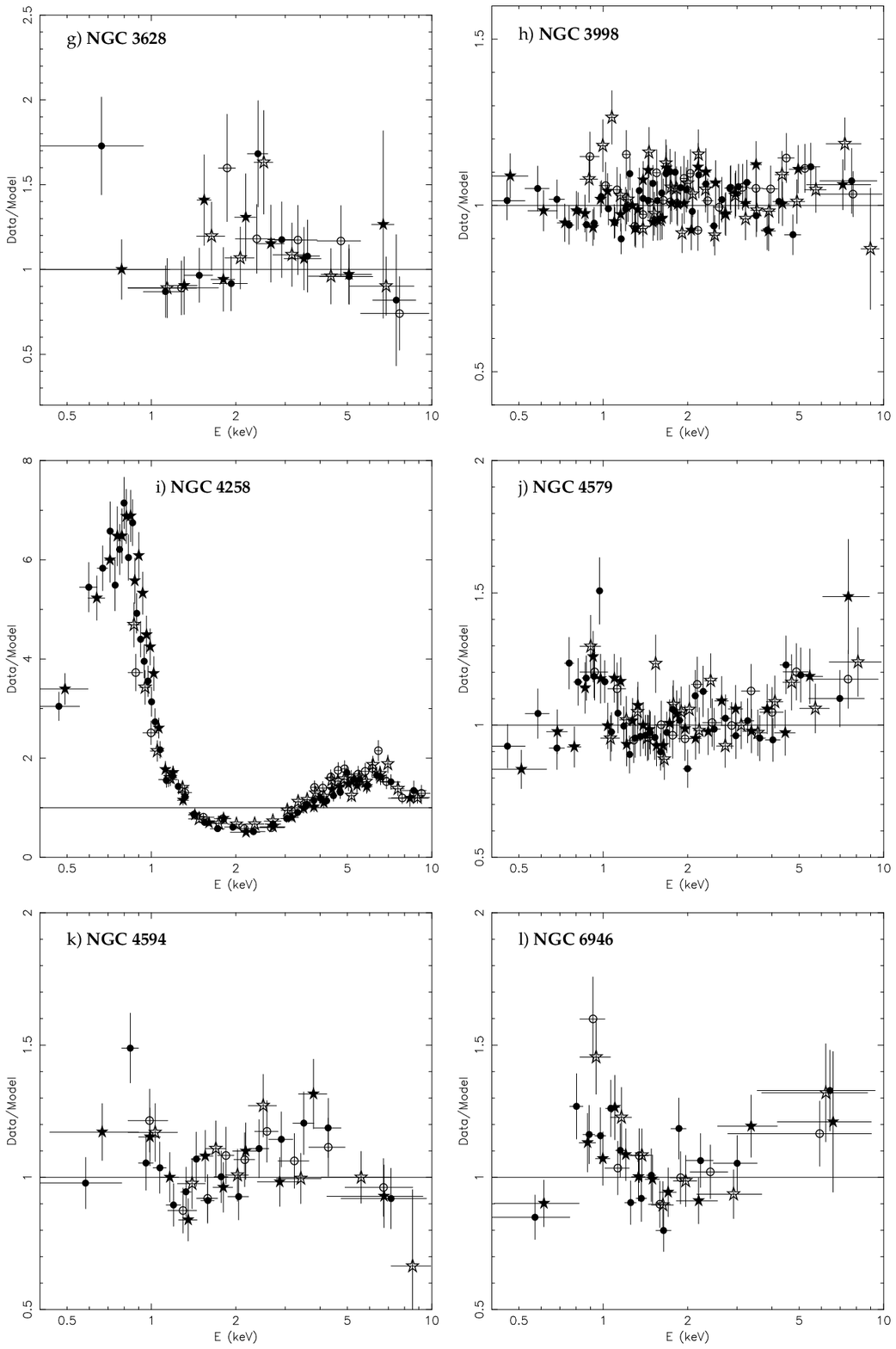}{7in}{0}{100.0}{100.0}{-350}{-170}
\caption{(cont.) Power-law Fit Data/Model Plots}
\end{figure*}
\setcounter{figure}{0}
\begin{figure*}
\plotfiddle{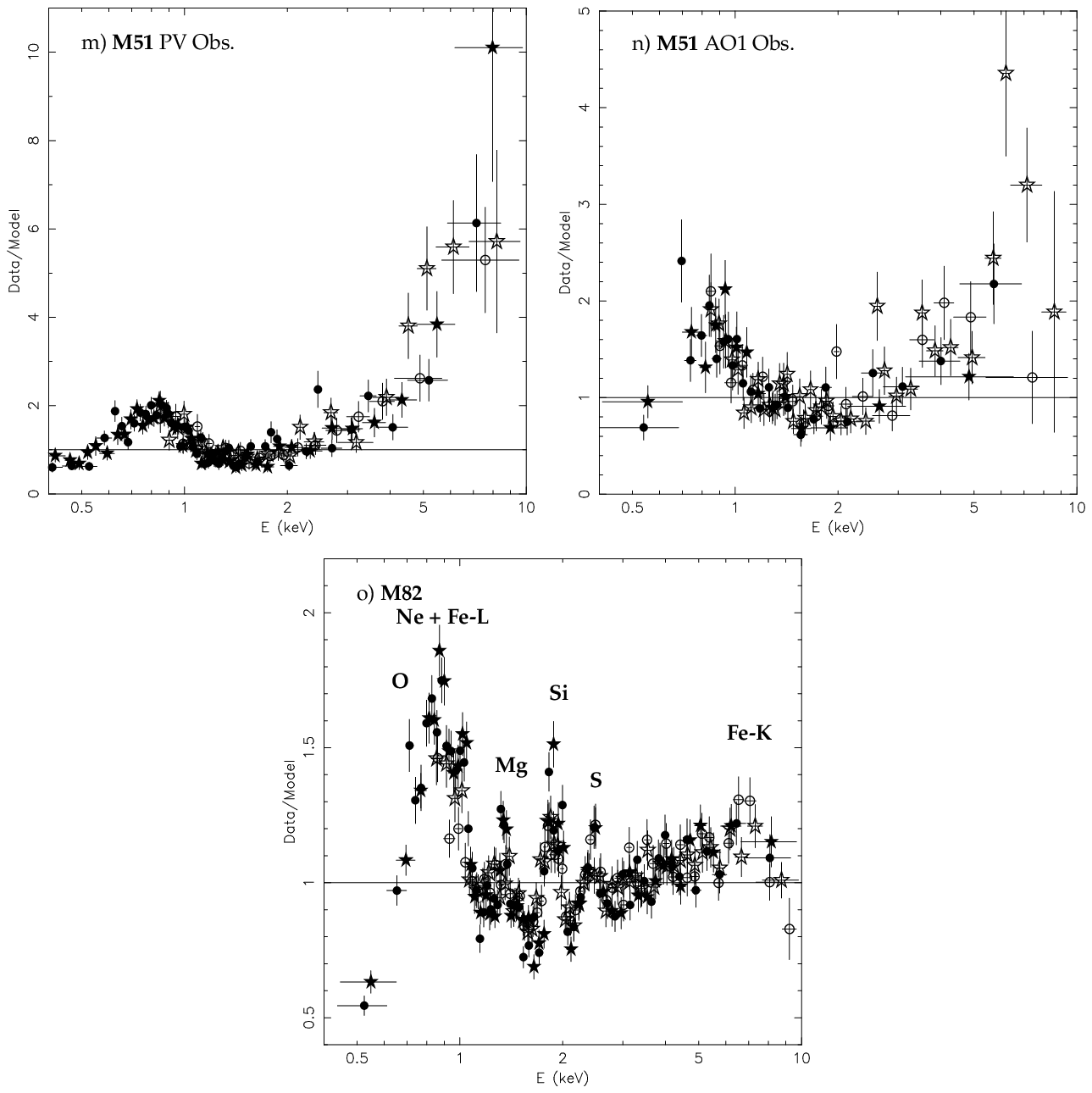}{7in}{0}{100.0}{100.0}{-350}{-170}
\caption{(cont.) Power-law Fit Data/Model Plots}
\end{figure*}
The simplest model that can be fit to the spectra is a power-law 
($N(E) = e^{-N_{H}\sigma(E)}E^{-\Gamma}$, where $\sigma(E)$ is the neutral 
material cross section and $N(E)$ is gives photons $\rm cm^{-2} \ s^{-1}
\ keV^{-1}$). The results of these fits are shown in Table 4.  For eight
 of the galaxies the fits were ``acceptable'' (i.e., with  
$\chi^{2}_{\nu} < 1.5$).  Ratios of data to best-fitting model for the
power-law fits are shown in Figure 1 where it can be seen
that significant residuals remain.  Significant line-like residuals
are evident in the cases of M82 and NGC 253, while in lower signal-to-noise
spectra residuals around 1 keV tend to be present.  These residuals 
imply the presence of an optically-thin plasma.  Accordingly, the
Raymond-Smith plasma model was also fit to the spectra
. In general there was no significant 
improvement in the fits, primarily becuase the temperature as 
determine by the centroid of the Fe-L complex centered around 1 keV 
tends to be less than 2 keV while the presence of significant emission
above 2 keV implies a harder spectrum.  In the cases of M51 and 
NGC 4258 it is clearly evident that more than one continuum component 
is required.  An important point is the fact that while the spectra of
some of the galaxies
such as NGC 3147 and NGC 3998 are fit well by an unabsorbed power-law model,
{\it none} of these
galaxies has an X-ray spectrum consistent with {\it only} 
an {\it absorbed} (i.e.,
$N_{H} \gtrsim 10^{21}\rm \ cm^{-2}$) power-law as might be expected from the
hypothesis that the X-ray emission of these galaxies is due solely to
the presence of an obscured AGN, without any other contribution to the X-ray
flux.

\subsection{Two-Component Fits}
\footnotesize
\begin{deluxetable}{crr}
\setcounter{table}{5}
\tablewidth{30pc}
\tablecaption{Raymond-Smith Plus Power-law and Double Raymond-Smith
Mean Fit Parameters}
\tablehead{
\colhead{Parameter} & \colhead{$\mu (\sigma_{\mu})$\tablenotemark{a}}
& \colhead{$\mu (\sigma_{\mu})$\tablenotemark{b}}
}
\startdata
$N_H \ (\times \ 10^{20}  \ \rm cm^{-2})$ & 4.9 (4.9) & 4.6 (4.0) \nl
$kT$ (keV) & 0.70 (0.10) & 0.69 (0.08) \nl
$A (/A_{\odot})$ & 0.043 (0.028) & 0.043 (0.028) \nl
$N_{H, 2} \ (\times \ 10^{22} \ \rm cm^{-2})$\tablenotemark{c}
& 1.2 (1.4) & 1.4 (2.1) \nl
$\Gamma$ or $kT_{2}$ (keV) & 1.71 (0.41) & 5.3 (3.2) \nl
$A_{2} \ (/A_{\odot})$ & \nodata & 0.19 (0.18) \nl
\tablenotetext{a}{Statistically-weighted mean of Raymond-Smith plus 
power-law fit parameters, with statisically-weight standard deviation 
given in parenthesis}
\tablenotetext{b}{As in $^a$, but for double Raymond-Smith fit}
\tablenotetext{c}{Absorption applied only to hard (i.e., power-law or 
second Raymond-Smith) component}
\enddata
\end{deluxetable}

\normalsize

\begin{figure*}
\plotfiddle{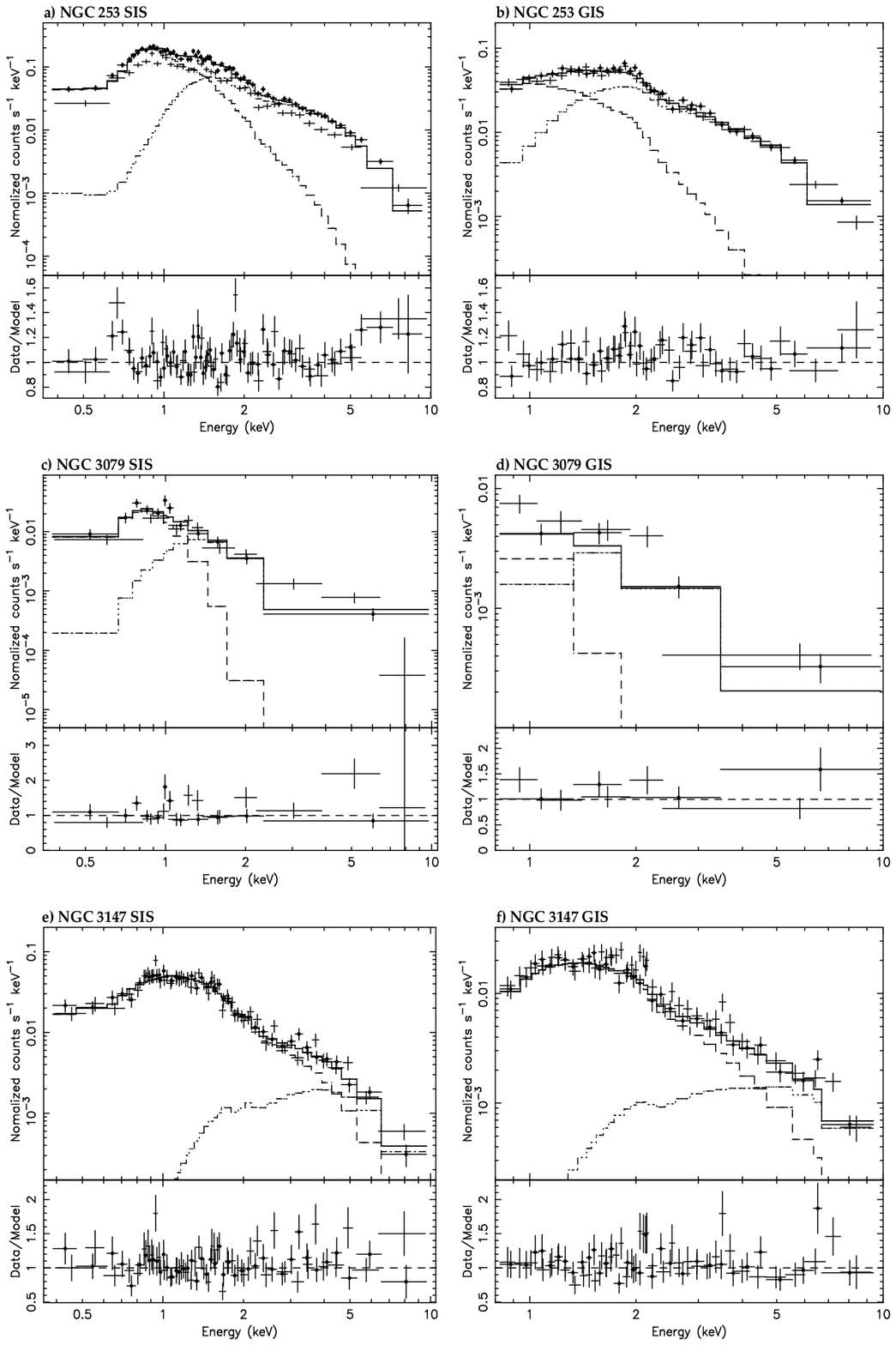}{7in}{0}{100.0}{100.0}{-350}{-170}
\caption{Best-fitting model, data, and data/model ratio for Raymond-Smith plus
Power-law fits to the {\it ASCA} data.  The best-fitting model is shown for
the S0 data only, for clarity.  The Raymond-Smith component is plotted with
a dashed line and the power-law component is plotted with a dot-dashed line.
The S0 and S2 points are marked (with filled circles) in the SIS and GIS
plots, respectively.}
\end{figure*}
\setcounter{figure}{1}
\begin{figure*}
\plotfiddle{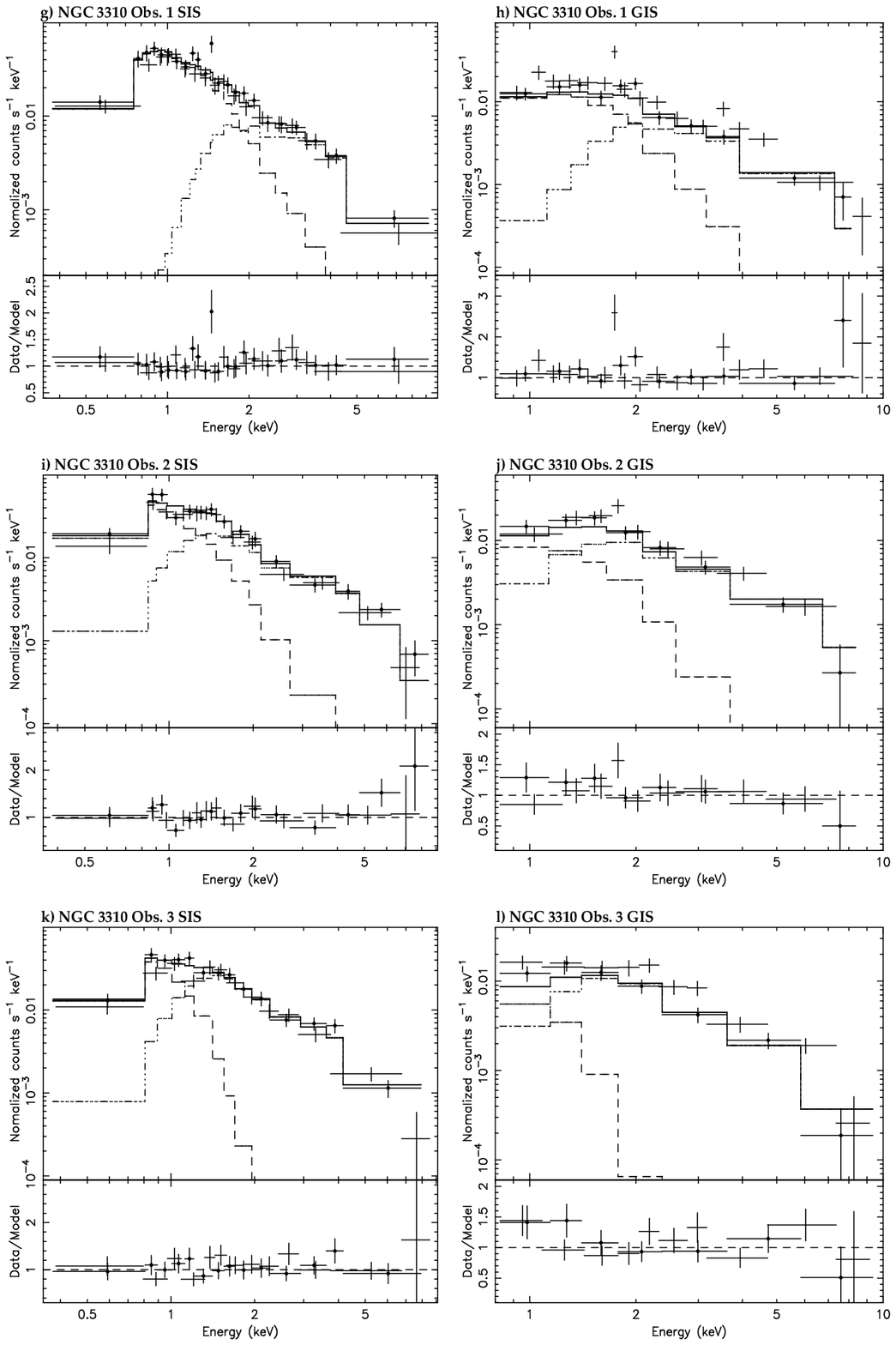}{7in}{0}{100.0}{100.0}{-350}{-170}
\caption{(cont.) Raymond-Smith plus Power-law Fits}
\end{figure*}
\setcounter{figure}{1}
\begin{figure*}
\plotfiddle{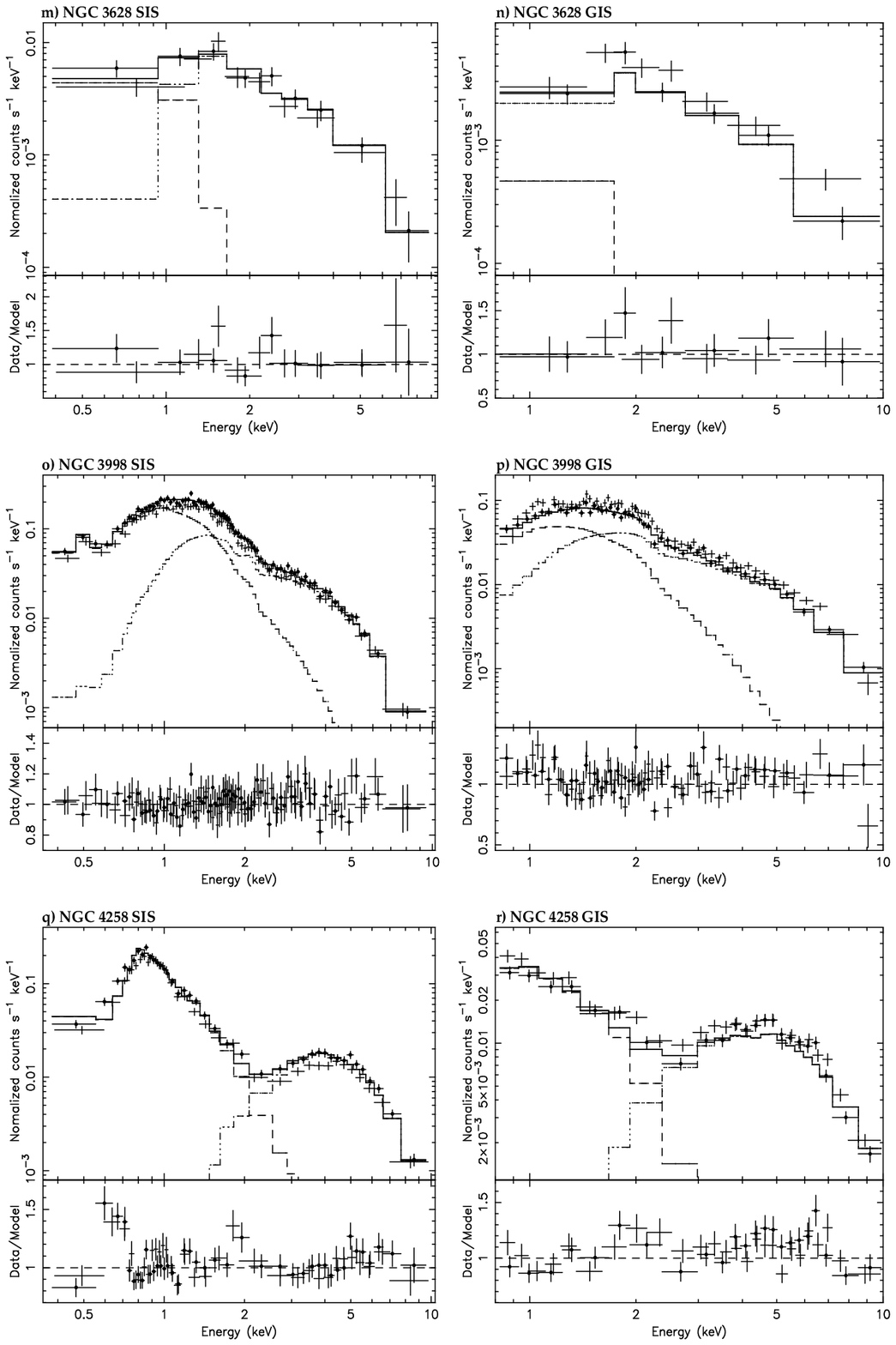}{7in}{0}{100.0}{100.0}{-350}{-170}
\caption{(cont.) Raymond-Smith plus Power-law Fits}
\end{figure*}
\setcounter{figure}{1}
\begin{figure*}
\plotfiddle{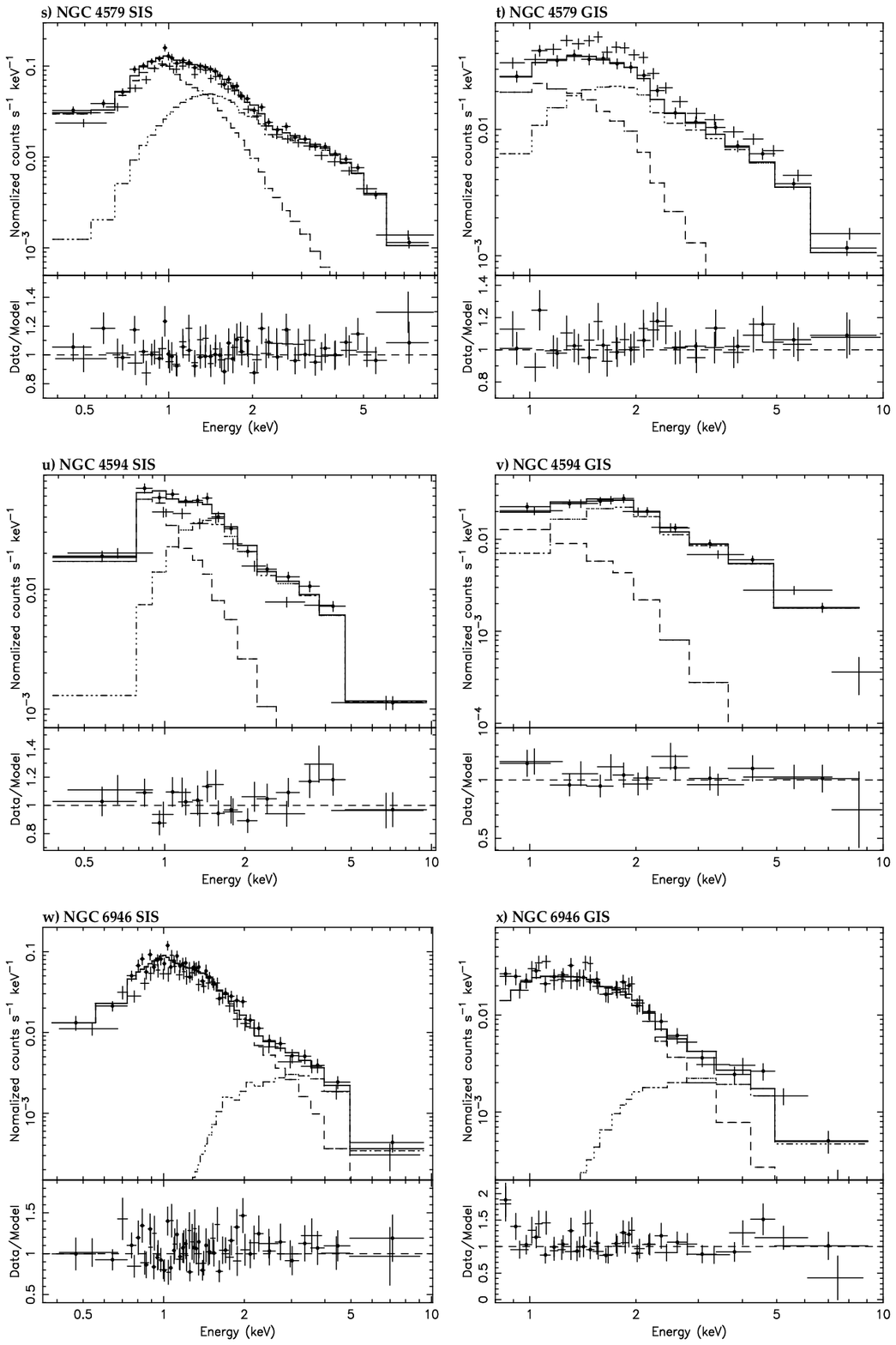}{7in}{0}{100.0}{100.0}{-350}{-170}
\caption{(cont.) Raymond-Smith plus Power-law Fits}
\end{figure*}
\setcounter{figure}{1}
\begin{figure*}
\plotfiddle{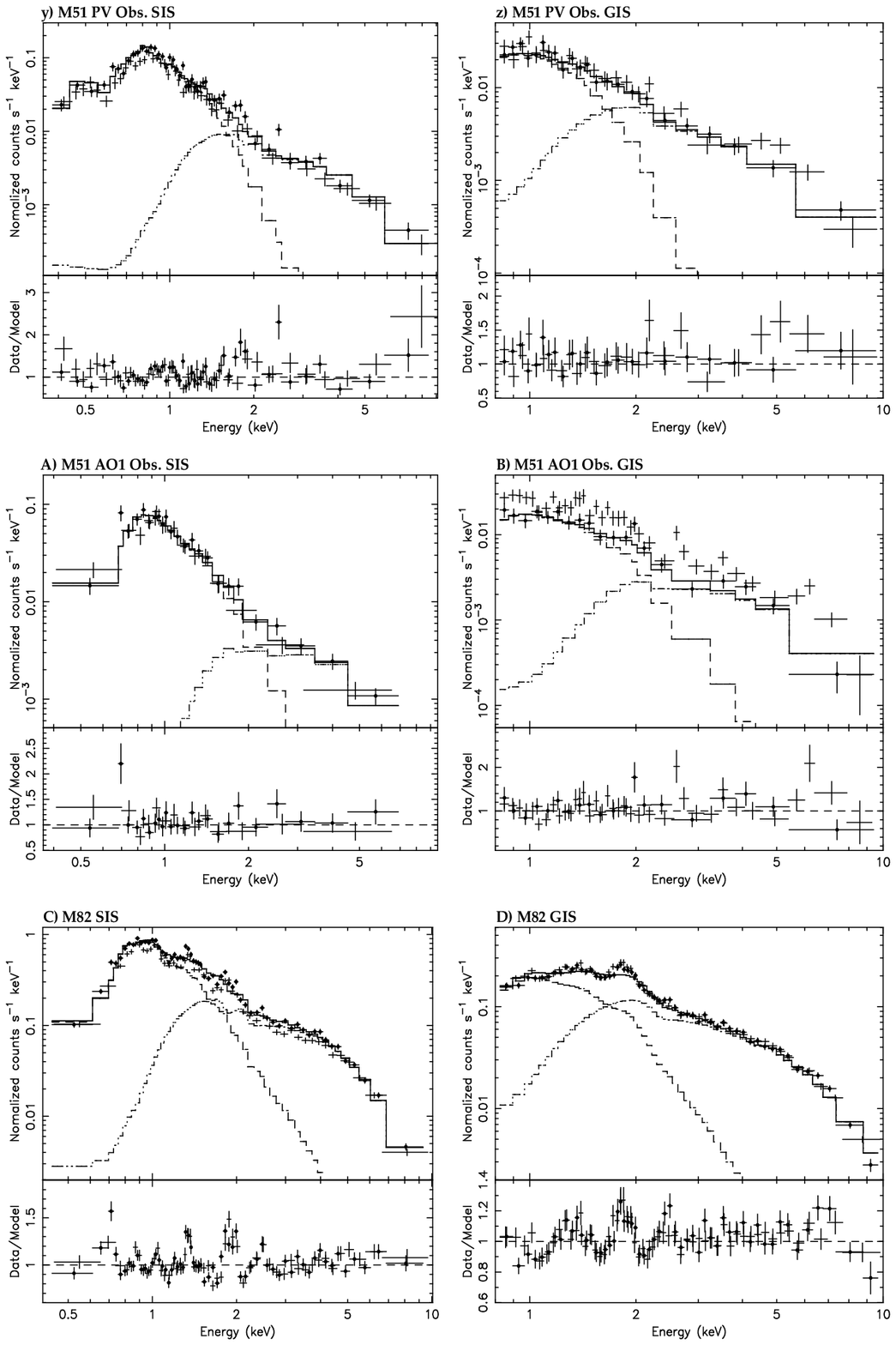}{7in}{0}{100.0}{100.0}{-350}{-170}
\caption{(cont.) Raymond-Smith plus Power-law Fits}
\end{figure*}
\begin{figure*}
\plotfiddle{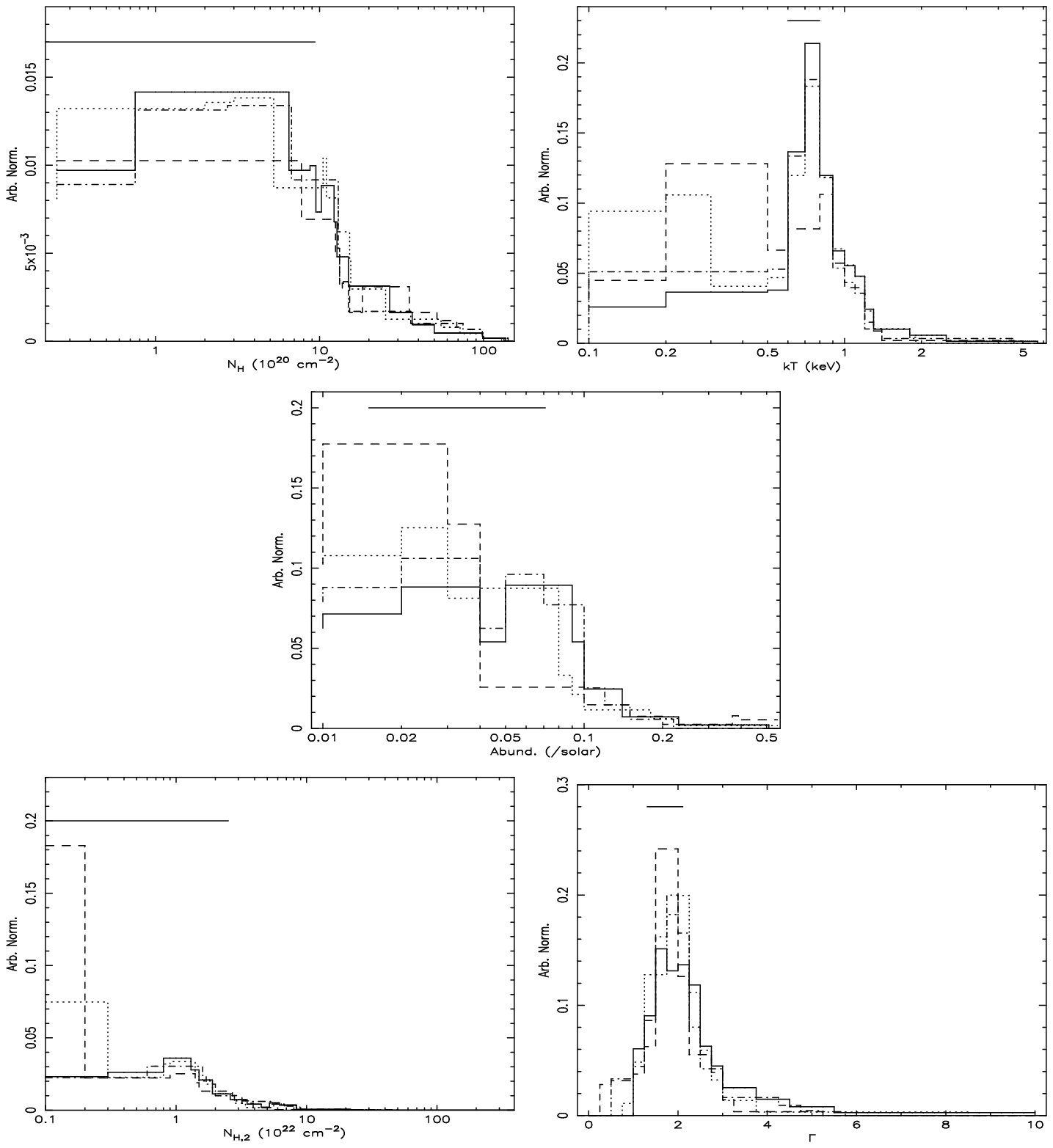}{5in}{0}{100.0}{100.0}{-340}{-260}
\caption{``Weighted'' histograms of the best-fitting
Raymond-Smith plus Power-law model fit parameters.  See text for details.}
\end{figure*}
 
In order to simultaneously fit the line-like features and hard
continuum, a two-component model consisting of a ``soft''
Raymond-Smith plasma component and a ``hard'' power-law component.  The 
results of fitting this model to the {\it ASCA} spectra are given
in Table 5.  In these fits the overall abundance was allowed to vary
while individual abundances were fixed at solar ratios.  Figures 
showing the model fits with residuals are given in Figure 2.  ``Weighted''
histograms for the model parameters are shown in Figure 3.  In these
histograms, the area occupied by each data point is held constant
while the width is proportional to the statistical error of the 
parameter.  In this way poorly-constrained parameters do not bias the
histograms significantly.  Evidently, the soft-component temperature
is rather narrowly distributed near $kT \sim 0.7$ keV.  Table 6 gives 
the statistically-weighted mean and standard deviation of each fit 
parameter, also showing the narrow distribution of soft-component 
temperature.  Also note that the mean value of the hard-component slope
is $\sim 1.7$, similar to the values of $\sim 1.7-2.0$ observed in 
``normal'' Seyfert 1 galaxies (c.f., \cite{mdp93}).

In order to check whether the X-ray spectra could be fit just as
well by any two-component model, double power-law fits were also
attempted.  However, in every case, the resultant $\chi^{2}$ was 
significantly higher, implying that the Raymond-Smith plus power-law 
model is preferred at better than 99\% confidence (for all of the S0-3, local
background fits with the exception of NGC 3147, where two components are
not required).  Additionally,
a double Raymond-Smith fit was attempted.  Here the improvement
in $\chi^{2}$ over the Raymond-Smith plus power-law fits (with one
additional parameter, namely the hard-component abundance) was 
significant only in the cases of M82 and NGC 4258.  This is 
principally due to the fact that in these cases there is significant
line emission near the ionized Fe-K line energies of 6.7-6.9 keV (see
also \S 6.4).  In
other cases no significant line emission is present near 6.7 keV, and
the signal-to-noise ratio is not sufficient to distinguish a 
non-thermal from a thermal continuum model for the hard (i.e., 2-10 
keV) component.  ``Weighted'' histograms of the fits parameters are 
shown in Figure 4.
\begin{figure*}
\plotfiddle{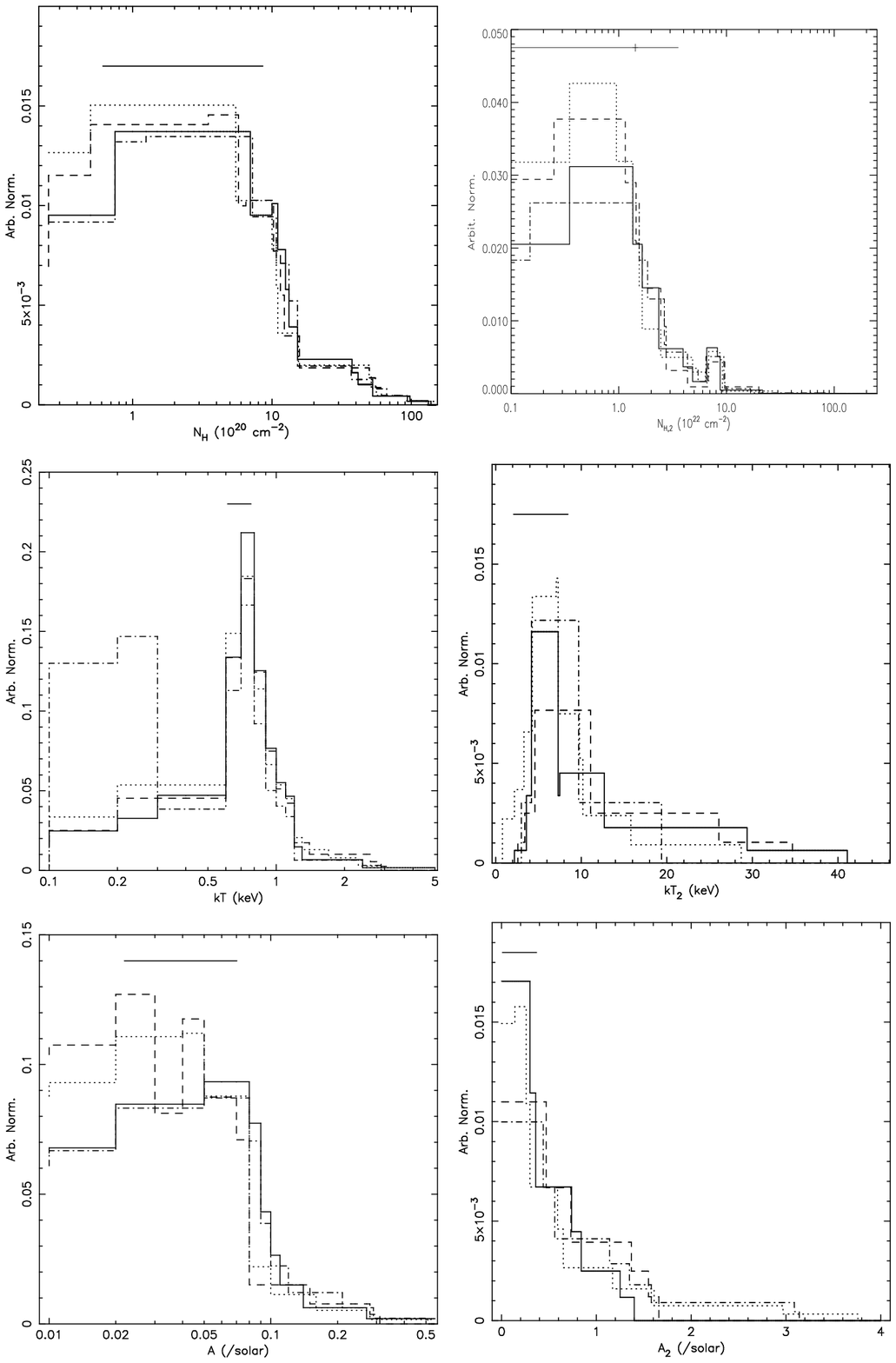}{7.5in}{0}{100.0}{100.0}{-340}{-130}
\caption{As in figure 3 but for the double Raymond-Smith fits.}
\end{figure*}

Table 6 also shows the statistically-weighted mean and 
standard deviation of the double 
Raymond-Smith fits.  Evidently, the soft component parameters do not 
depend strongly on choice of hard-component model. This is shown 
graphically in Figure 5 which shows correlations of the common 
parameters (i.e., $N_{H}$, $kT$ and $A$ (from the soft-component in 
the case of the double Raymond-Smith fits), and the (additional) 
absorption applied to the hard component.  However, note that the 
soft-component abundances may nevertheless have large systematic errors
since there is 
no reason to assume {\it a priori} that the soft-component is {\it 
entirely} thermal.  Any addition contribution to the 0.5-2.0 
continuum that is not explicitly taken into account in these fits 
would act to reduce the {\it observed} equivalent widths and hence 
the abundance.  This is one possible explanation for the unusually 
low abundances observed in these galaxies since starburst activity is 
likely to enrich the interstellar medium with the heavy metals that 
dominate X-ray spectra (see below for a more detailed discussion of 
the abundances).  In addition, it is likely that there are multiple 
temperatures (and/or a temperature gradient) present in any X-ray 
emitting gas.  However, since the signal-to-noise in most of the 
galaxies in this sample do not warrant fits with more than two 
components, this possibility can only be examined in detail in a few 
objects (e.g., M82 in \cite{ptak97}, \cite{tsuru97}, and Moran 
\& Lehnert 1997).
\begin{figure*}
\plotfiddle{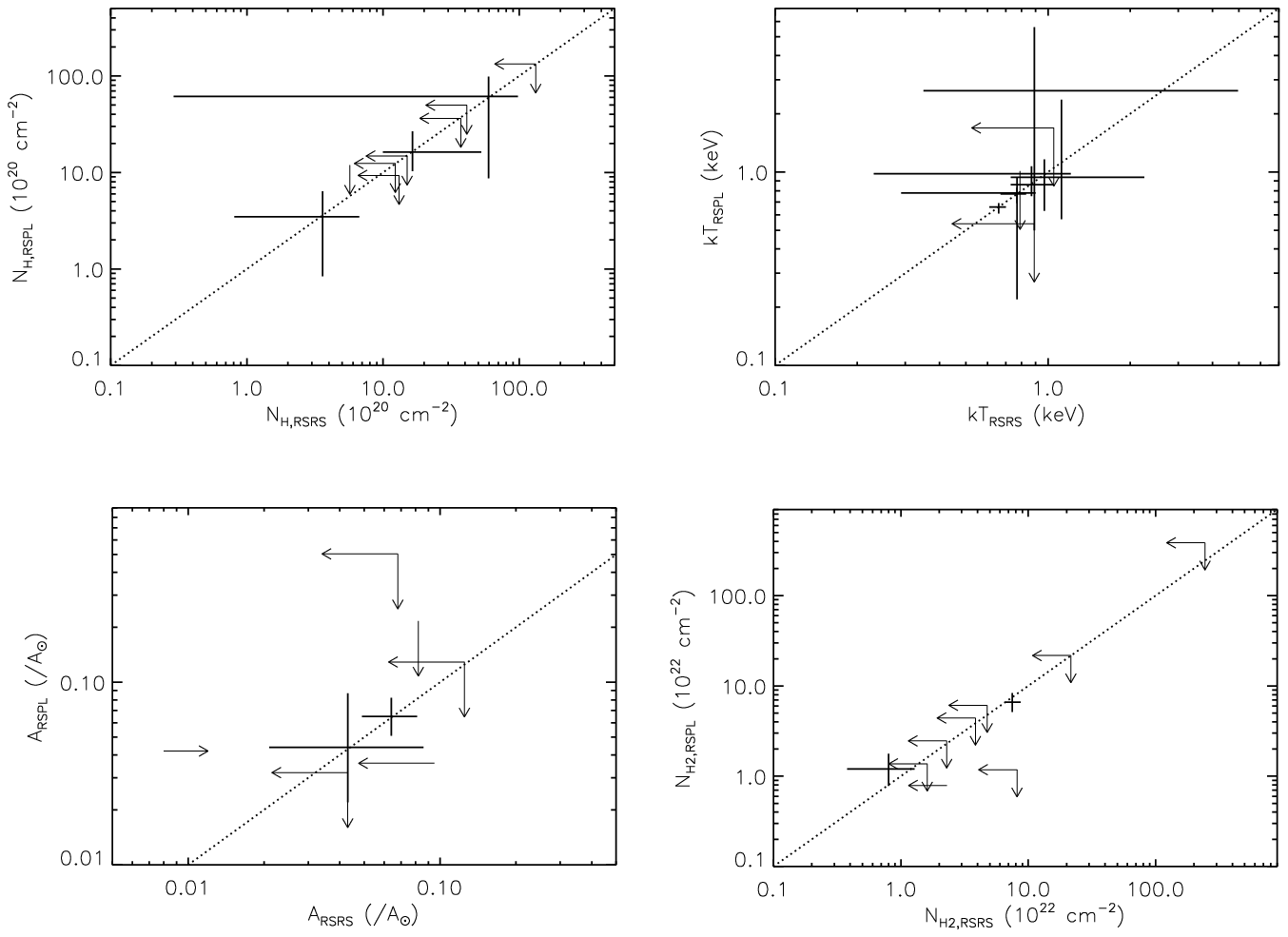}{5in}{0}{100.0}{100.0}{-330}{-150}
\caption{Comparison of the best-fit values for $N_H$, kT, A
(abundance), and $N_{H,2}$ (absorption applied only to the power-law
component, in excess of the $N_H$ column density applied to the entire model)
from the Raymond-Smith plus power-law (``RSPL'') and double Raymond-Smith fits
(``RSRS'').}
\end{figure*}

In the cases of NGC 253, NGC 4258, M51 and M82, it is clear from Figure 2 
that significant sharp spectral features remain below $\sim$ 2.5 keV implying 
that modeling the soft flux with a single plasma with relative abundances 
(i.e., Fe/O) at the solar ratios is clearly too simple.  Since the 
strongest line emission expected is Fe-L emission 
around 1 keV, it is likely that the overall abundance determination is 
dominated by the Fe abundance inferred from the Fe-L complex.  Accordingly, 
for these galaxies additional fits were attempted in which the Fe abundance 
was allowed to vary independently from the other elements.  The results are 
shown in Table 7 where the Fe abundance relative to the 
other heavy elements (dominated by the $\alpha$-burning elements O, Mg,
S and Si) tend to be lower than the solar ratio by factors of 2-4. A 
similar low relative abundance of Fe was observed in more detailed 
spectral fitting of M82 and NGC 253 (Ptak et al. 1997) where other 
abundance ratios were allowed vary, and in Tsuru et al. (1997) where 
three-component fits were attempted to M82. In 
the case of these galaxies, the Fe abundance is depressed relative to the 
$\alpha$-process elemental abundances by factors of 5 or more, although
as discussed in \S6.7, this effect is diminished when a
multiple-temperature model is fit to the data.

Allowing the oxygen abundance to be free results in a 
statistically significant improvement in $\chi^{2}$ in NGC 4258 and 
M51, which are ``Seyfert 2'' galaxies, but {\it not} in M82 and NGC 253,
which are ``starburst'' galaxies.  While these fits are probably too 
simplistic to be taken too literally (i.e., the fits can be rejected
statistically in all cases), {\it this result nevertheless 
implies a quantifiable difference in the starburst-like spectra of 
in Seyfert 2 and pure starburst galaxies}. 
We note that it would be unlikely
for calibration errors to affect starbursts and Seyfert 2 galaxies
differently, given the similarities in the overall shape of their spectrum.
However, in order to determine to what extent these results may be biased by
the 
SIS calibration uncertainties, the NGC 4258 and M51 SIS-only fits were
repeated only using data in the 0.6-10.0 keV bandpass.  In the case of NGC
4258, the best-fit oxygen abundance dropped from $\sim 0.47$ to $\sim 0.24$,
with the other fit parameters not changing signicantly.   The change in
$\chi^2$ for allowing the oxygen abundance to vary dropped from 23.0 to 4.3.
In the case of M51, the oxygen abundance dropped to 0.02, while the Fe
and remaining abundances were 0.025 and 0.094, respectively, similar to the
values found for the fits in which the oxygen abundance
was tied with the other abundances, and
the change in $\chi^2$ remained at $\sim 8.6$.  Thus, this result is evidently
more robust in the case of M51.  To check the statistical significance of the
overabundance of oxygen in the case of NGC 4258, we simulated source and
background SIS spectra using the best-fitting model with the oxygen abundance
left free, and repeated the fitting process.  We found that the oxygen
abundance was found to be three or more times the Fe abundance $\sim 90\%$ of
the time, and two or more times the ``other'' abundances $\sim 68\% \
(1\sigma)$ of the time.  Incidentally, these simulations also showed that the
statistical errors on the fits with the oxygen and iron abundance left free
are approximately 1/2 of the values shown in Table 7, indicating that the
confidence limits given in this paper are conservative.
\subsection{{\it ROSAT} PSPC and {\it ASCA} Fits}
While the {\it ROSAT} PSPC has limited spectral resolution, as 
mentioned above its 
sensitivity in the 0.1-0.3 keV range and the presence of the large 
carbon edge results in a higher level of accuracy in determining absorption 
columns on the order of $10^{19-20} \rm cm^{-2}$.  Accordingly,  
simultaneous fits were attempted with {\it ASCA} and PSPC spectra.  
Separate PSPC observations existed in several cases (see Table 2) and 
these were not combined as an attempt to observe any long-term variability.
For each galaxy two sets of PSPC spectra were extracted - one with the 
same region size as the {\it ASCA} SIS source region and one with a 
1.25' source region.  A 1.25' region is appropriate for a 
point-source observation (in the case of off-axis PSPC observation 
``ancillary'' responses were created to compensate for the loss of 
flux due to vignetting and blurring of the point-spread function).  
Hereafter the 1.25' spectra are denoted ``nuclear'' since they are 
dominated by nuclear flux, while the ``{\it ASCA}''-sized source regions 
should result in roughly the same flux as the {\it ASCA} fits (unless the 
galaxy's X-ray flux is significantly extended beyond several 
arcminutes or the X-ray flux varied between the {\it ASCA} and {\it ROSAT}
observations).  The results are shown in Table 8 (the results of the 
S0-3 only, ``local'' background fits from Table 5 are shown for 
comparison) and Figure 6.  As expected, the addition of the PSPC 
spectra resulted in significantly smaller errors on the column 
density while the other fit parameters were not impacted 
significantly.  Note that systematic offsets on the order of $\sim 15\%$
between the {\it ASCA} 
SIS and PSPC are apparent in the residuals shown in Figure 6.  In most 
cases the statistical errors dominate this systematic error.  These offsets
are probably dominated by error in the calibration of the PSPC, since the
are present present above 0.6 keV (the SIS calibration is most suspect
below 0.6 keV, however see \S6.7), where the SIS calibration is on the order of 6\%
(\cite{orr98}; also note that version 4.0 of {\sc ftools} was used in this
analysis, which underestimated spectral corrections to PSPC spectra).
The 
discrepancy is particularly acute in the case of NGC 3998, and it is 
possible that the column density varied between the {\it ROSAT} and 
{\it ASCA} observations in this case.  Table 9 gives the ratio of the
0.5-2.0 keV fluxes inferred from the PSPC spectra relative to the 
{\it ASCA} spectra.  The variation in the flux inferred from each {\it ASCA}
detector is shown to give an indication of the accuracy of the {\it 
ASCA} fluxes.  There are clear variations present among the ASCA and 
{\it ROSAT} fluxes (particularly in the cases of NGC 3628 and NGC 4579) and 
among the ``nuclear'' and ``{\it ASCA}''-region PSPC fluxes (indicating a 
0.5-2.0 keV spatial extent in excess of 1.25').
\begin{figure*}
\plotfiddle{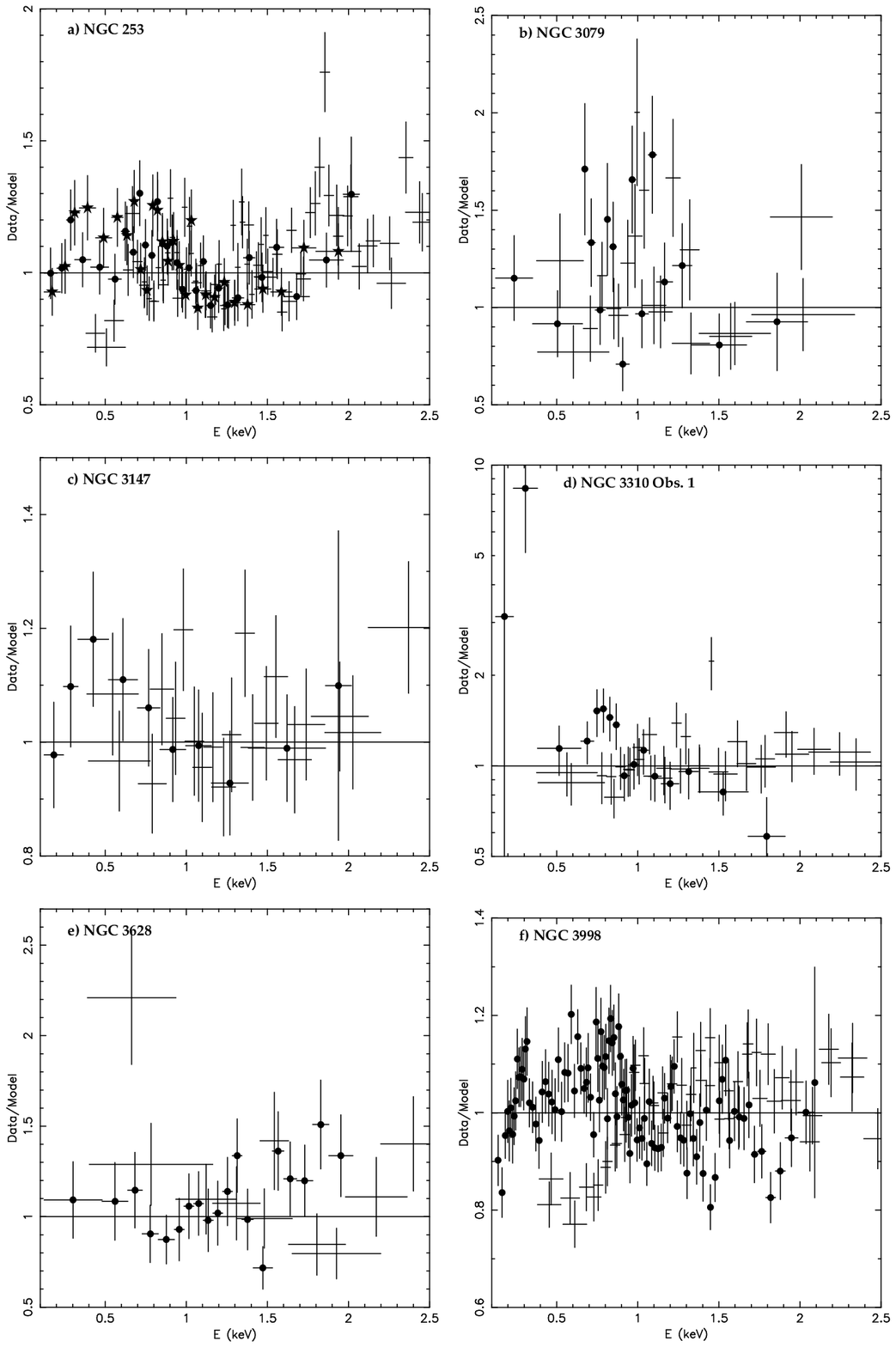}{7in}{0}{100.0}{100.0}{-320}{-150}
\caption{Ratio of data to model for Raymond-Smith
plus Power-law fits to {\it ASCA} and {\it ROSAT} PSPC spectra.  Only the
energy range where {\it ASCA} and {\it ROSAT} overlap is shown.  The PSPC
points are marked with filled circles (in cases where more than one PSPC data
set is available, the additional data points are marked with filled stars).}
\end{figure*}
\setcounter{figure}{5}
\begin{figure*}
\plotfiddle{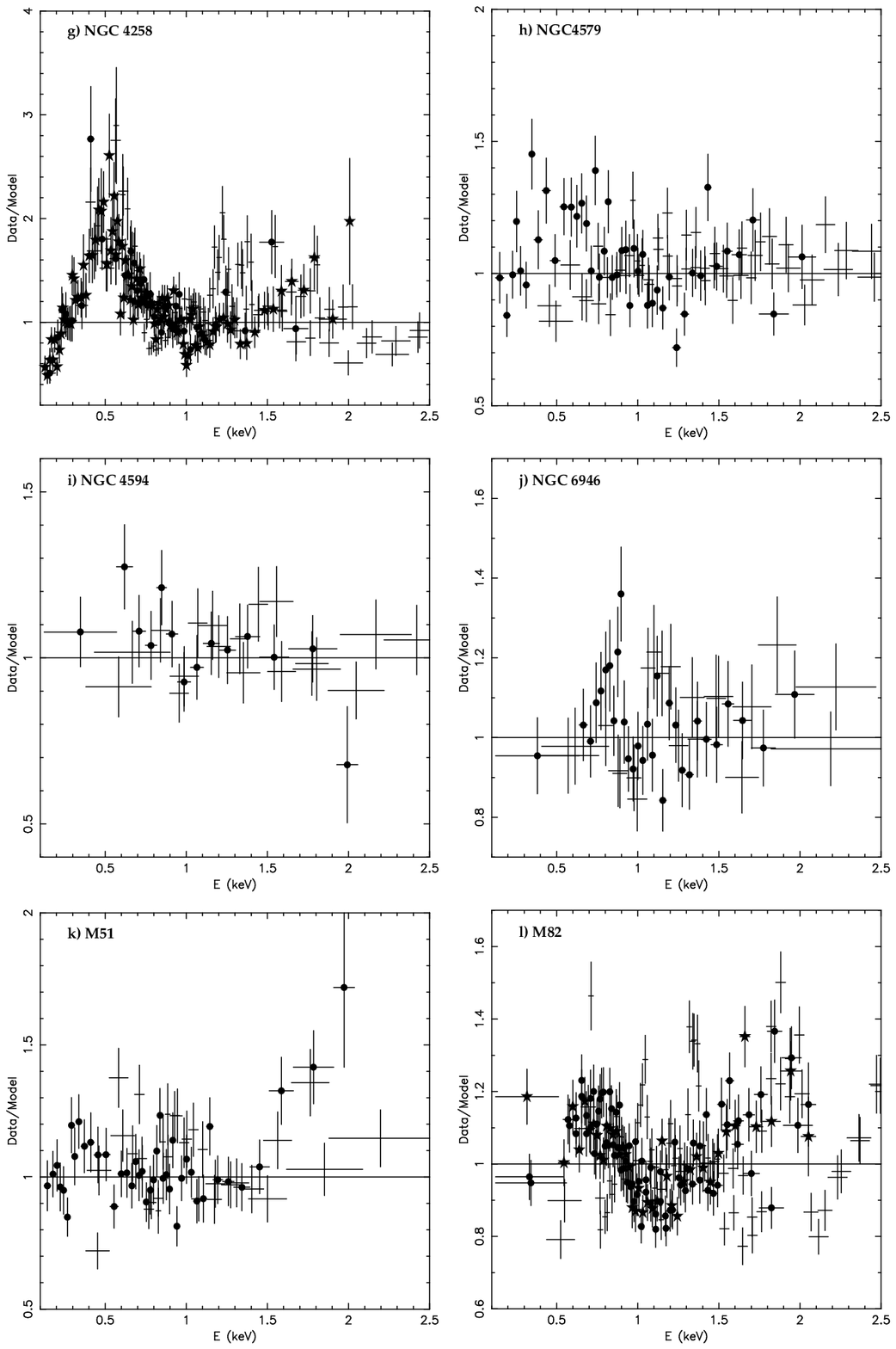}{7in}{0}{100.0}{100.0}{-350}{-150}
\caption{(cont.) {\it ASCA} and {\it ROSAT} Raymond-Smith plus Power-Law Fits}
\end{figure*}

Since the addition of the PSPC data did not change the best-fit temperature
determination, it is highly unlikely that the clustering of temperatures around
the mean of 0.7 keV is due simply to a bandpass effect (i.e., the PSPC data is
extending the bandpass from 0.4 keV to 0.1 keV).  In galaxies where the
best-fit temperature does change, either the Raymond-Smith component is not
very significant (e.g., NGC 3147) or it appears that the true variation is in
the best fit $N_H$, and the temperature is changing due to a correlation
between temperature and $N_H$.  Note that the change in $N_H$ in these cases
may not necessarily be real, but rather is indicative of flux below 0.4 keV
that is detected by {\it ROSAT}.  Indeed, in every case where $N_H$ changes
with the addition of PSPC data, $N_H$ decreases.  This flux is most likely
due to halo gas at $\sim 0.3$ keV which is also occasionally dominates the
{\it ASCA}-only fits (c.f., NGC 3079).  Another indication that the clustering
around the temperature of 0.7 keV (and secondarily around 0.3 keV) is not
a bandpass effect is that in most cases the best-fit temperatures are fairly
well constrained, which would not be the case otherwise.

\subsection{Fe-K Emission}
In order to search for Fe-K emission, commonly observed in ``normal'' 
Seyfert galaxies, the {\it ASCA} spectra were fit in the 3-10 keV 
range with a simple power-law where the hard-component dominates.  
Initialy a narrow Gaussian was introduced, fixed at 6.4 keV 
(appropriate for Fe less ionized the Fe XVI) and at 6.7 keV 
(appropriate for He-like Fe, the ionization state expected at the 
temperatures implied by the hard component of the double Raymond-Smith 
fits).  In the cases where the Gaussian significantly reduced 
$\chi^{2}$ the centroid and physical width ($\sigma$) of the Gaussian
was allowed to be a free 
parameter.  The results are shown in Table 10 (which give 90\% 
confidence upper-limits for galaxies with no significant Fe-K 
emission).  Significant Fe-K emission was detected in five galaxies: 
M51 (see also \cite{T98a}), M82 (see also \cite{ptak97}),
NGC 3147 (see also \cite{ptak96}), NGC 4579 (see also 
\cite{T98b}), and
NGC 4258 (see also \cite{mak94}).  See Paper II for a 
discussion of the implications of the presence or lack of Fe-K emission
in these galaxies.

\subsection{Ionized Absorbers}
\footnotesize
\setcounter{table}{10}
\begin{deluxetable}{llll}
\tablewidth{30pc}
\tablecaption{Ionized Oxygen Absorption Edge Search}
\tablehead{
\colhead{Galaxy} & \colhead{Fit\tablenotemark{*}}
& \colhead{$\tau_{\rm OVII}$ (0.74 keV)}
& \colhead{$\tau_{\rm OVIII}$ (0.87 keV)}}
\startdata
NGC 3147 &  S0-1, PL & $<$0.20 & $<$0.13 \nl
&S0-3, PL & $<$0.16 & $<$0.10 \nl
NGC 3998 & S0-1, PL & 0.13 (0.006-0.26) & 0.08 ($<$0.20) \nl
&S0-3, PL & 0.07 ($<$0.19) & 0.01 ($<$0.11) \nl
NGC 4579 & S0-1, PL & $<$0.04 & $<$0.04 \nl
&S0-1, RS + PL & 0.14 ($<$1.78) & $<$0.62 \nl
\tablenotetext{*}{Fit used in oxygen edge search; PL = power-law-only 
model, RS + PL = Raymond-Smith + power-law model (both models include 
neutral absorption as discussed in \S4.1-2).}
\enddata
\end{deluxetable}
\normalsize
Ionized, or ``warm,'' absorbers are frequently observed in Seyfert 1s as evidenced 
by the presence of O VII and OVIII absorption edges at 0.74 and 0.87 keV 
(see Reynolds 1997; George et al. 1997).  Two good candidates for 
searching for a warm absorber in LINERs are NGC 3998 and NGC 4579.  Both 
are broad-line LINERs, making them analogous to Seyfert 1s, and are among the 
brightest of the galaxies observed in this thesis.  While also relatively 
bright, the soft flux in NGC 253, NGC 4258, M51 and M82 is dominated by 
thermal flux (note that this flux is often spatially as well as spectrally 
distinct from the hard flux).  Table 11 shows the 
upper-limits on OVII and OVIII edges for these two galaxies obtained using 
the simple power-law fits (still including the neutral absorption).  Since 
the presence of a soft component is significant in NGC 4579, the 
upper-limits were re-determined using the Raymond-Smith + power-law model 
discussed in \S 6.2, which resulted in far less restrictive limits 
on the edges than the simple power-law model fits (as would be expected 
since the presences of the plasma flux ``fills-in'' the edges and reduces the 
contrast with the underlying continuum).  Also, in Seyfert 
2s where the X-ray emission is thought to be dominated by scattered flux, 
the scattering medium must be highly ionized.  Here also absorption due to 
highly ionized oxygen might be observable.  Accordingly, limits on OVII and 
OVIII were also determined for NGC 3147 (again using the power-law only 
model since the presence of a soft component in not highly significant).

\subsection{Fluxes and Luminosities}
The observed and intrinsic 
0.4-10.0 keV fluxes for the {\it ASCA} spectra are given in Table 12.  
The S0-3, local background Raymond-Smith plus power-law fits and
{\it ASCA} and {\it ROSAT} PSPC (with {\it ASCA} SIS source regions) 
simultaneous fits were used to determine the 
fluxes.  The intrinsic fluxes were determined by setting $N_{H}$ and 
$N_{H,2}$ 
to zero.  As stated above, the values of $N_{H}$ determined from the 
{\it ASCA} 
and PSPC simultaneous fits are more reliable than the values of $N_{H}$
determined from the fits with {\it ASCA} data alone, so the unabsorbed
fluxes from these fits are 
likewise more reliable.  The contributions to the 0.4-10.0 keV flux 
from the Raymond-Smith and power-law components individually is also 
given.  The significance of the improvement to $\chi^{2}$ of the 
Raymond-Smith plus power-law fit over the power-law fit to the spectra 
from NGC 3147 is not large so the fluxes determined from the 
power-law fit is shown.  Also, the flux from NGC 3310 does not extend 
beyond 1.25' and the ``nuclear'' PSPC spectrum has better 
signal-to-noise than the SIS-region PSPC spectrum, so the nuclear 
PSPC spectrum fit is also used in this case to determined the fluxes.  
This example also demonstrates how large the systematic error due to 
the specific choice of spectral model can be when estimating the 
unabsorbed flux of each component.  The total observed 0.4-10.0 keV luminosity
is also given in Table 12.

\section{Complex Models}
Here we investigate the results of fitting more complicated models to the
spectra from 
brightest galaxies in our sample, NGC 253, NGC 4258, M51 and M82.
Specifically, it is physically probable that the gas in these galaxies is not
isothermal but rather has some temperature structure.
Table 13 gives the results 
of double Raymond-Smith plus power-law fits.  As discussed in Ptak et
al. (1997) and DWM, the abundances inferred from spatially-averaged are
somewhat model-dependent.  Evidently, modeling the soft flux with even a
crude temperature
structure results in absolute abundances that are no longer significantly
subsolar, and in most cases, the underabundance of Fe relative to
$\alpha$-process elements is no longer significant.  For a detailed discussion
of this issue, see DWM.  Table 13 shows the results of restricting the SIS
bandpass to 0.6-10.0 keV, and, with the exception of the NGC 253 fits, the
results are comparable, indicating that these results, along with the
simpler fits discussed previously, are not aversely affected by SIS
calibration uncertainties below 0.6 keV.

\footnotesize
\setcounter{table}{13}
\begin{deluxetable}{lll}
\tablewidth{15cm}
\tablecaption{Power-law Temperature Plasma plus Power-Law Model Fit to M82}
\tablehead{
\colhead{Fit Parameter} & \colhead{SIS Data Only} &  \colhead{SIS + PSPC Data}
}
\startdata
$N_H  \ (10^{21}  \ \rm cm^{-2})$ & 2.9 (2.0-3.2) & 1.1 (0.9-1.4) \nl
kT$_max$ (keV) & 1.10 (0.71-1.73) & 1.12 (0.70-1.72) \nl
$\alpha$ & 0.01 ($<$0.41) & 0.44 (0.25-0.60) \nl
$A \ (/A_{\odot})$ & 0.12 ($<$0.32) & 0.49 ($<$2.4) \nl
$A \ (/A_{\odot,O})$ & 0.14 (0.05-0.24) & 1.1 (0.9-3.0) \nl
$A \ (/A_{\odot,Mg})$ & 0.42 (0.25-0.62) & 4.1 (0.9-9.6) \nl
$A \ (/A_{\odot,Si})$ & 0.55 (0.38-0.73) & 5.8 ($>$1.2)\nl
$A \ (/A_{\odot,S})$ & 0.57 ($>$0.20) & 4.5 ($>$2.0)\nl
$A \ (/A_{\odot,Fe})$ & 0.12 (0.09-0.14) & 1.2 (0.3-1.8) \nl
$N_{H, 2}  \ (10^{22}  \ \rm cm^{-2})$ & 1.7 ($<$2.9) & 0.34 (0.08-0.62) \nl
$\Gamma$ & 1.62 (1.55-1.73) & 1.47 (1.33-1.55) \nl 
$\chi^2$/dof & 479.5/384 ($>$99.9\%) & 1271/938 ($>$99.9\%) \nl
\enddata
\tablecomments{The soft-component model is a plasma consisting of a
superposition of power-law distribution of temperatures.  In practice,
the model is computed discretely with the weighting of each temperature $T$
being proportional to $(\frac{T}{T_{max}})^{\alpha}$.  Elemental abundances
that were allowed to vary freely of the others are specified.}
\end{deluxetable}
\normalsize
Table 14 gives the results of fitting the {\it ASCA}
 and
PSPC spectra from M82 with a plasma model with a power-law temperature
dependence (a power-law continuum component is still included as well to fit
the hard flux). 
However, we note that the temperatures inferred from
a (relatively) simple Raymond-Smith plus power-law fit {\it does} adequately
represent the average temperature of the X-ray emitting gas, and it is
therefore of interest to discuss the thermodynamics implied by the
Raymond-Smith plus power-law fits (see paper II).  For example, we used
the best-fitting model listed in Table 14 to simulate an SIS observation
of M82, and fit the simulation with a single temperature plasma plus
power-law model (with the Fe
abundance allow to vary).  This resulted
in a temperature of 0.66 keV, similar to the value listed
in Table 7, where a single temperature plasma plus power-law model was fit of
the actual SIS observation of M82.  This is an important point
for faint galaxies where more complex fits are not statistically warranted.

%
\section{Summary}
We have presented the results of a systematic analysis of the {\it ASCA} and
{\it ROSAT} PSPC X-ray spectra from a sample of low-luminosity AGN, LINERs
and starbursts galaxies.  The presence of at least two components appears to
be universal for these types of galaxies.  The ``soft'' component is
well-described by a thermal plasma model with $T \sim 10^7$ K while the
``hard'' component is well-described by an absorbed power-law.  The hard
component may also be due to a thermal bremsstrahlung component with $T \sim
10^{8}$ K, however the soft component is inconsistent with a featureless
continuum (i.e., a power-law or zero-abundance plasma model) at a high level
of confidence.  The relative strength of the hard and soft component varies
from galaxy to galaxy, with the intrinsic 0.4-10.0 keV luminosity of the
soft component typically being in the range of $10^{39-40} \ \rm ergs \
s^{-1}$ and the hard component typically being in the range of
$10^{40-41} \ \rm ergs \ s^{-1}$.
The absolute abundances inferred from the soft-component fits
tend to be significantly sub-solar, although the absolute abundances may be
uncertain.  In the galaxies with X-ray emission bright enough for individual
abundances to be observed, the relative abundance of Fe to $\alpha$-process
elements tend to be low (see also \cite{ptak97}).  There is some indication
(at a low-statisical signficance) that the abundance properties starburst
emission from starburst galaxies differs from the starburst emission from
low-luminosity AGN.
Ionized absorption is not
observed in these galaxies, in contrast to typical Seyfert 1 galaxies,
although absorption edges may be ``washed-out'' by the soft component.
Fe-K emission is detected in several galaxies.  See Ptak (1997)
and Ptak et al. (1998a,b) for a detailed discussion of these results.
It is likely that detailed analysis of these galaxies by telescopes capable
of (better) spatially-resolved broadband spectroscopy (i.e., {\it AXAF} and
{\it XMM}) will considerably improve our understanding of the X-ray emission
from these galaxies. 

\acknowledgements
A.P. wishes to thank the NASA GSRP program for support. This research made
extensive use of the HEASARC (at NASA/GSFC), NASA's Astrophysics Data System
Abstract Service, and NED (at IPAC).

\end{document}